\documentclass[12pt]{article}
\usepackage{epsfig}
\usepackage{axodraw}
\newcommand{\Bbar}{\bar B}
\newcommand{\cbar}{\bar c}
\begin{document}
%%%%%%%%%%%%%%%%%%%%%%%%%%%%%%%%%%%%%%%%%%%%%%%%%%%%%%%%%%%%%%%%%%%%%%
% equations
\def\bfix{~\newline\centerline{XXXXXXXXXXX corrected versionXXXXXXXXXXXXXX}\newline}
\def\efix{~\newline\centerline{XXXXXXXXXXXX corrected version ends hereXXXX}\newline}
\def\bfix{}
\def\efix{}
\def\bXX{}
\def\eXX{}
\def\ket{{\rangle}}
\def\bra{{\langle}}
\def\ie{{\it i.e.}}
\def\be{\begin{equation}}
\def\ee{\end{equation}}
\def\ba{\begin{eqnarray}}
\def\ea{\end{eqnarray}}
\def\bq{\begin{quotation}\noindent}
\def\eq{\end{quotation}}
\def\mref#1{Eq. (\ref{Eq:#1})}
\def\mreff#1{Fig. \ref{Eq:#1}}
\def\mreft#1{Table \ref{Eq:#1}}
\def\mlab#1{\label{Eq:#1}}
\def\mlabf#1{\label{Eq:#1}}
\def\mlabt#1{\label{Eq:#1}}
\def\half{\frac{1}{2}}
\def\to{\rightarrow}
\def\nn{\nonumber\\}
\def\sk{\vskip 1cm}
\def\skk{\vskip 3mm}
\def\mat#1#2#3{\langle{#1}\vert{#2}\vert{#3}\rangle}
\def\etal{{\it et al.}}
\def\etc{{\it etc.~}}

%%%%%%%%%%%%%%%%%%%%%%%%%%%%%%%%%
\def\ibid#1#2#3{{\it ibid. }{\bf #1} #2 {(#3)}}
\def\PR#1#2#3 {{\it Phys. Rev. }{\bf D#1} #2 {(#3)}}
\def\PRL#1#2#3 {{\it Phys. Rev. Lett. }{\bf #1} #2 {(#3)}}
\def\PL#1#2#3 {{\it Phys. Lett. }{\bf #1} #2 {(#3)}}
\def\AP#1#2#3 {{\it Ann, Phys. }{\bf #1} #2 {(#3)}}
\def\ZP#1#2#3 {{\it Z. Phys. }{\bf #1} #2 {(#3)}}
\def\NP#1#2#3 {{\it Nucl. Phys. }{\bf #1} #2 {(#3)}}
\def\MPL#1#2#3 {{\it Mod. Phys. Lett. }{\bf #1} #2 {(#3)}}
\def\NC#1#2#3 {{\it Nuov. Cim. }{\bf #1} #2 {(#3)}}
\def\PREP#1#2#3 {{\it Phys. Report }{\bf #1} #2 {(#3)}}
\def\PROG#1#2#3 {{\it Prog. Theor. Phys. }{\bf #1} #2 {(#3)}}
\def\SOV#1#2#3{{\it Sov. J. Nucl. Phys. }{\bf #1} #2 {(#3)}}
\def\JETP#1#2#3{{\it JETP }{\bf #1} #2 {(#3)}}
\def\RMP#1#2#3{{\it Rev. Mod. Phys. }{\bf #1} #2 {(#3)}}
%matrix element

%%%%%%%%%%%%%%%%%%%%%%%%%%%%%%%%%%%%%%%%%%%%%%%%%%%%%%%%%%%%%%%%%%%%%
\def\phiout{{\phi_a^{out}}}
\def\phiin{\phi_a^{in}}
\def\psiout{\psi_a^{out}}
\def\psiin{\psi_a^{in}}
\def\vin{\phi_{a\mu}^{in}}
\def\vout{\phi_{a\mu}^{out}}
\def\ofx{{(x)}}
\def\op{{\bf P}}
\def\oc{{\bf C}}
\def\ot{{\bf T}}
\def\cp{{\bf CP}}
\def\cpt{{\bf CPT}}
\def\vecr{{\vec r}}
\def\vecn{{\vec {\nabla}}}
\def\vecA{{\vec A}}
\def\psibar{\overline\psi}
\def\outin{{{out}\choose{in}}}
\def\inout{{{in}\choose{out}}}

\def\vr{\vec x}
\def\itt{{\it T}}
\def\b{{\bf b}}
\def\a{{\bf a}}
\def\d{{\bf d}}

\def\alphadot{{\dot\alpha}}
\def\betadot{{\dot\beta}}
\def\gammadot{{\dot\gamma}}

%%%%%%%%%%%%%%%% took out c^2
\def\N{\sqrt{{{E+m}\over{2m}}}}
\def\F#1{{{#1}\over{E+m}}}
%%%%%%%%%%%%%%%%%
\def\itemm{\hangindent\parindent\textindent}
\def\noi{\noindent}
\def\onehead#1{\vskip1pc\leftline{\bf #1}}
\def\twohead#1{\vskip1pc\leftline{\bf #1}}
\def\ts{\thinspace}

\def\sq2{{1\over{\sqrt{2}}}}
\def\omegaar{{\vec{\omega}}}
\def\kbar{\overline K}
\def\Pbar{\overline P}
\def\Abar{\overline A}
\def\dbar{\overline d}
\def\ubar{\overline u}
\def\sbar{\overline s}
\def\tbar{\overline t}
\def\nubar{\overline \nu}
\def\cbar{\overline c}
\def\Dbar{\overline D}
\def\Bbar{\overline B}
\def\Kbar{\overline K}

\def\bbar{\overline b}
\def\gbar{\overline g}
\def\pbar{\overline p}
\def\qbar{\overline q}
\def\vecr{\vec r}

\def\dm{\Delta m}
\def\ss{(1+s^2)}
\def\cdmp{cos\Delta m(t_1+t_2)}
\def\cdm{cos\Delta m(t_1-t_2)}

\def\g5{\gamma_5}
\def\gm{(1-\gamma_5)}
\def\gp{(1+\gamma_5)}

% vectors
\def\mvec#1{\vec{#1}\,}
\def\pslash{\mbox{/\llap p}}
\def\slash#1{\mbox{/\llap #1}}

% small roman indices
\def\msmall#1{\mbox{\rm \small #1}}
%%%%%%%%%%%%%%%%%%%%%%%%%%%%%%%%%%%%%%%%%%%%%%%%%
%bigi
%\renewcommand{\topfraction}{1.0}
%\renewcommand{\bottomfraction}{1.0}
%\renewcommand{\textfraction}{0.0}
\newcommand{\matel}[3]{\langle #1|#2|#3\rangle}
\newcommand{\hscale}{\mu\ind{hadr}}
\newcommand{\aver}[1]{\langle #1\rangle} 
\renewcommand{\Im}{\mbox{Im}\,}
\renewcommand{\Re}{\mbox{Re}\,}
\newcommand{\GeV}{\,\mbox{GeV}}
\newcommand{\MeV}{\,\mbox{MeV}}
\newcommand{\BR}{\,\mbox{BR}}
\newcommand{\dd}{{\rm d}}
\def\d{{\bf d}}
%%%%%%%%%%%%%%%%%%%%%%%%%%%%%%%%%%%%%%%%%%%%%%%%%%%%

\vskip 0.5cm
%%%%%%%%%%%%%%%%%%%%%%%%%%%%%%%%%%%%%%%%%%%%%%%%%%%%%%%%%%%%%
%%%%%%%%%%%%%%%%%%%%%%%%%%%%%%%%%%%%%%%%%%%%%%%%%%%
\renewcommand{\thefootnote}{\fnsymbol{footnote}}
\newcommand{\cita} [1] {$^{\hbox{\scriptsize \cite{#1}}}$}
\newcommand{\prepr}[1] {\begin{flushright}  {\bf #1} \end{flushright}}
\newcommand{\titul}[1] {\begin{center}{\Large {\bf #1 } }\end{center}}

\newcommand{\autor}[1] {\begin{center}{\bf \lineskip .3cm #1} \end{center}}

\newcommand{\adress}[1] {\begin{center}  {\normalsize \bf \it #1 }
\end{center}}
\newcommand{\abstr}[1] {{\begin{center} \vskip .5cm {\bf \large Abstract
                        \vspace{0pt}} \end{center}}\begin{quote} \small #1
                        \end{quote}}
%%%%%%%%%%%%%%%%%%%%%%%%%%%%%%%%%%%%%%%%%%%%%%%%%%%%%%%%%%%%%%%%

\begin{titlepage}
\titul{\bf Nonfactorizable contributions to $B\to D^{(*)}M$ decays}

\vskip1.0cm

\autor{ Yong-Yeon Keum$^{a}$
\footnote{yykeum@eken.phys.nagoya-u.ac.jp},
T. Kurimoto$^{b}$\footnote{krmt@k2.sci.toyama-u.ac.jp},
Hsiang-nan Li$^{c,d}$\footnote{hnli@phys.sinica.edu.tw}, \\
Cai-Dian L\"u$^{e}$\footnote{lucd@ihep.ac.cn},
and A.I. Sanda$^{a}$\footnote{sanda@eken.phys.nagoya-u.ac.jp}}

\vskip1.5cm
\adress{$^{a}$Department of
Physics, Nagoya University, Nagoya, Japan}

\adress{$^{b}$Faculty of Science, Toyama University, Toyama
930-8555, Japan}

\adress{$^{c}$Institute of
Physics, Academia Sinica, Taipei, Taiwan 115, Republic of China}

\adress{$^{d}$Department of Physics, National
Cheng-Kung University, \\
Tainan, Taiwan 701, Republic of China}

\adress{$^{e}$CCAST (World
Laboratory), P.O. Box 8730, Beijing 100080, China;}

\adress{$^{e}$Institute of High Energy Physics, CAS, P.O. Box
918(4), Beijing 100039, China\footnote{Mailing address}}

%\newpage
\vskip 1.0cm
\begin{abstract}
While the naive factorization assumption works well for many
two-body nonleptonic $B$ meson decay modes, the recent measurement
of $\bar B\to D^{(*)0}M^0$ with $M=\pi$, $\rho$ and $\omega$ shows
large deviation from this assumption. We analyze the $B\to
D^{(*)}M$ decays in the perturbative QCD approach based on $k_T$
factorization theorem, in which both factorizable and
nonfactorizable contributions can be calculated in the same
framework. Our predictions for the Bauer-Stech-Wirbel parameters,
$\left|a_2/a_1\right|= 0.43\pm 0.04$ and $Arg(a_2/a_1)\sim
-42^\circ$ and $\left|a_2/a_1\right|= 0.47\pm 0.05$ and
$Arg(a_2/a_1)\sim -41^\circ$, are consistent with the observed
$B\to D\pi$ and $B\to D^*\pi$ branching ratios, respectively. It
is found that the large magnitude $|a_2|$ and the large relative
phase between $a_2$ and $a_1$ come from color-suppressed
nonfactorizable amplitudes. Our predictions for the ${\bar B}^0\to
D^{*0}\rho^0$, $D^{*0}\omega$ branching ratios can be confronted
with future experimental data.
\end{abstract}

\vskip2.0cm
%{\bf PACS number(s): 12.38.Bx, 13.35.Hw, 12.38.Qk, 11.10.Hi}
\end{titlepage}
%\newpage

\section{INTRODUCTION}

Understanding nonleptonic $B$ meson decays is crucial for testing
the standard model, and also for uncovering the trace of new
physics. The simplest case is two-body nonleptonic $B$ meson
decays, for which Bauer, Stech and Wirbel (BSW) proposed the naive
factorization assumption (FA) in their pioneering work \cite{BSW}.
Considerable progress, including generalized FA
\cite{Cheng94,Cheng96,Soares} and QCD-improved FA (QCDF)
\cite{BBNS}, has been made since this proposal. On the other hand,
technique to analyze hard exclusive hadronic scattering was
developed by Brodsky and Lepage \cite{LB} based on collinear
factorization theorem in perturbative QCD (PQCD). A modified
framework based on $k_T$ factorization theorem was then given in
\cite{BS,LS}, and extended to exclusive $B$ meson decays in
\cite{LY1,CL,YL,CLY}. The infrared finiteness and gauge invariance
of $k_T$ factorization theorem was shown explicitly in \cite{NL}.
Using this so-called PQCD approach, we have investigated dynamics
of nonleptonic $B$ meson decays \cite{KLS,LUY,KS}. Our
observations are summarized as follows:
\begin{enumerate}
\item FA holds approximately for charmless $B$ meson decays, as
our computation shows that nonfactorizable contributions are
always negligible due to the cancellation between a pair of
nonfactorizable diagrams.

\item Penguin amplitudes are enhanced, as the PQCD formalism
includes dynamics from the region, where the energy scale $\mu$
runs to $\sqrt{\bar\Lambda m_b}<m_b$, $\bar\Lambda\equiv
m_B-m_b$ being the $B$ meson and $b$ quark mass difference.

\item Annihilation diagrams contribute to large short-distance
strong phases through the $(S+P)(S-P)$ penguin operators.

\item The sign and the magnitude of CP asymmetries in
two-body nonleptonic $B$ meson decays can be calculated, and we
have predicted relatively large CP asymmetries in the $B\to
K^{(*)}\pi$
\cite{KLS,Keum02} and
$\pi\pi$ modes\cite{LUY,KS,Keum01}.
\end{enumerate}

All analyses involving strong dynamics suffer large theoretical
uncertainties than we would like. How reliable are these
predictions? This can be answered only by comparing more of our
predictions with experimental data. For this purpose, we study the
$B\to D^{(*)}M$ decays in PQCD, where $M$ is a pseudoscalar or a
vector meson. A $D^{(*)}$ meson is massive, and the energy release
involved in two-body charmed decays is not so large. If
predictions for these decays agree reasonably well with
experimental data, PQCD should be more convincing for two-body
charmless decays. Since penguin diagrams do not contribute, there
are less theoretical ambiguities, such as the argument on chiral
enhancement or dynamical enhancement. Checking the validity of
PQCD analyses in charmed decays is then more direct.

Note that FA is expected to break down for charmed nonleptonic $B$
meson decays \cite{NPe}. FA holds for charmless decays because of
the color transparency argument: contributions from the dominant
soft region cancel between the two nonfactorizable diagrams, where
the exchanged gluons attach the quark and the antiquark of the
light meson emitted from the weak vertex. For charmed decays with
the light meson replaced by a $D^{(*)}$ meson, the two
nonfactorizable amplitudes do not cancel due to the mass
difference between the two constituent quarks of the $D^{(*)}$
meson. Hence, nonfactorizable contributions ought to be important.
This observation further leads to the speculation that strong
phases in the $B\to D^{(*)}M$ decays, if there are any, arise from
nonfactorizable amplitudes. In charmless decays, strong phases
come from annihilation amplitudes through the $(S+P)(S-P)$ penguin
operators, since nonfactorizable ones are negligible as explained
above. Annihilation amplitudes should not be the source of strong 
phases for charmed decays, which do not involve the
$(S+P)(S-P)$ penguin operators. 

In this paper we shall apply the PQCD formalism to the two-body
charmed decays $B\to D^{(*)}M$ with $M=\pi$, $\rho$ and $\omega$.
PQCD has predicted the strong phases from annihilation amplitudes for
charmless decays, which are consistent with the recently measured
CP asymmetries in the $B^0_d\to\pi^+\pi^-$ modes. It is
then interesting to examine whether PQCD also gives the correct
magnitude and strong phases from nonfactorizable amplitudes implied 
by the isospin relation of the $B\to D^{(*)}M$ decays.
Compared to the work in \cite{YL}, the contributions from the
twist-3 light meson distribution amplitudes and the threshold 
resummation effect have been taken into account, and more modes 
analyzed. The power counting rules for charmed $B$ meson decays, 
constructed in \cite{TLS2}, are employed to obtain the leading 
factorization formulas. It will be shown that nonfactorizable 
contributions to charmed decays are calculable in PQCD, and play an 
important role in explaining the isospin relation indicated by 
experimental data. The predictions for the $B\to D^{*0}\rho^0$, 
$D^{*0}\omega$ branching ratios can be confronted with future 
measurement.

In Sec.~II we review the progresses on the study of
two-body charmed nonleptonic $B$ meson decays in the literature.
The PQCD analysis of the above decays is presented in Sec.~III
by taking the $B\to D\pi$ modes as an example. Numerical results
for all the $B\to D^{(*)}M$ branching ratios, and for the extracted
BSW parameters $a_1$ and $a_2$ are collected in Sec.~IV. In
Sec.~V we compare the PQCD approach to exclusive
$B$ meson decays with others in the literature. Sec.~VI
is the conclusion. Appendix A contains the explicit expressions of
the factorization formulas.

\section{REVIEW OF PREVIOUS WORKS}

\subsection{PQCD Approach to $B\to D^{(*)}$ Form Factors}

To develop the PQCD formalism for charmed $B$ meson decays, we have
investigated the $B\to D^{(*)}$ transition form factors in the
large recoil region of the $D^{(*)}$ meson \cite{TLS2}. We briefly
review this formalism, which serves as the
basis of the $B\to D^{(*)}M$ analysis. The $B\to D^{(*)}$
transition is more complicated than the $B\to\pi$ one, because it
involves three scales: the $B$ meson mass $m_B$, the $D^{(*)}$
meson mass $m_{D^{(*)}}$, and the heavy meson and heavy quark mass
difference, $\bar\Lambda=m_B-m_b\sim m_{D^{(*)}}-m_c$ of order of
the QCD scale $\Lambda_{\rm QCD}$, $m_{D^{(*)}}$ ($m_c$) being the
$D^{(*)}$ meson ($c$ quark) mass. We have postulated the hierachy
of the three scales,
\begin{eqnarray}
m_B\gg m_{D^{(*)}}\gg\bar\Lambda\;,
\label{hie}
\end{eqnarray}
which allows a consistent power expansion in $m_{D^{(*)}}/m_B$ and
in $\bar\Lambda/m_{D^{(*)}}$.

\begin{figure}
\begin{center}
\epsfig{file=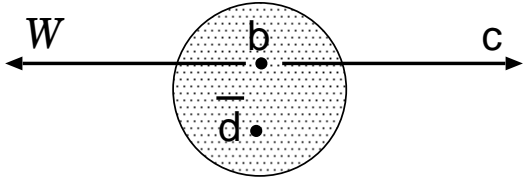,width=7cm}
\end{center}
\caption{$B\to D^{(*)}$ transition through the $b$ quark decay into a
$c$ quark and a virtual $W$ boson.}
\label{leading}
\end{figure}

Write the $B$ ($D^{(*)}$) meson momentum $P_1$ ($P_2$) in the
light-cone coordinates as
\begin{eqnarray}
P_1=\frac{m_B}{\sqrt{2}}(1,1,{\bf 0}_T)\;,\;\;\;\;
P_2=\frac{m_B}{\sqrt{2}}\left(1,\frac{m_{D^{(*)}}^2}{m_B^2},
{\bf 0}_T\right)\;.
\end{eqnarray}
The picture associated with the $B\to D^{(*)}$ transition is shown 
in Fig.~\ref{leading}, where the initial state is approximated by 
the $b\bar d$ component. The $b$ quark decays into a $c$ quark and 
a virtual $W$ boson, which carries the momentum $q$. Since the
constituents are roughly on the mass shell, we have the invariant
masses $k_i^2\sim O(\bar\Lambda^2)$, $i=1$ and 2, where $k_1$ ($k_2$)
is the momentum of the spectator $\bar d$ quark in the $B$ ($D^{(*)}$)
meson. The above kinematic constraints lead to the order of 
magnitude of $k_1$ and $k_2$ \cite{TLS2},
\begin{eqnarray}
k_1^\mu&\sim& (\bar\Lambda,\bar\Lambda,\bar\Lambda)\;,
\nonumber\\
k_2^\mu&\sim& \left(\frac{m_B}{m_{D^{(*)}}}\bar\Lambda,
\frac{m_{D^{(*)}}}{m_B}\bar\Lambda,\bar\Lambda\right)\;.
\label{k2}
\end{eqnarray}

\begin{figure}
\begin{center}
\epsfig{file=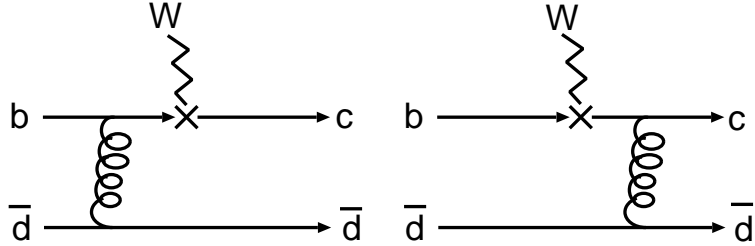,width=10cm}
\end{center}
\caption{Lowest-order diagrams contributing to the $B\to D^{(*)}$
form factors. Quite a bit of momentum must be transfered to the
spectator $\bar d$ quark through the hard gluon exchange.}
\label{gluon}
\end{figure}

The lowest-order diagrams contributing to the $B\to D^{(*)}$ form
factors contain a hard gluon exchange between the $b$ or $c$ quark
and the $\bar d$ quark as shown in Fig.~\ref{gluon}. The $\bar d$
quark undergoes scattering in order to catch up with the $c$
quark, forming a $D^{(*)}$ meson. With the parton momenta in
Eq.~(\ref{k2}), the exchanged gluon is off-shell by
\begin{eqnarray}
(k_1-k_2)^2\sim -\frac{m_B}{m_{D^{(*)}}}\bar\Lambda^2\;,
\label{hs1}
\end{eqnarray}
which has been identified as the characteristic scale of the hard
kernels. Under Eq.~(\ref{hie}), we have $m_B/m_D^{(*)}\gg 1$, and
the hard kernels are calculable in perturbation theory. It has been found
that the applicability of PQCD to the $B\to D^{(*)}$ transition at
large recoil is marginal for the physical masses $m_B$ and
$m_{D^{(*)}}$ \cite{TLS2}.

Infrared divergences arise from higher-order corrections to
Fig.~\ref{gluon}. The soft (collinear) type of divergences is
absorbed into the $B$ ($D^{(*)}$) meson wave function
$\phi_B(x_1,b_1)$ ($\phi_{D^{(*)}}(x_2,b_2)$), which is
not calculable but universal. The impact
parameter $b_1$ ($b_2$) is conjugate to the transverse momentum
$k_{1T}$ ($k_{2T}$) carried by the $\bar d$ quark in the $B$
($D^{(*)}$) meson. It has been shown, from equations of motion for
the relevant nonlocal matrix elements, that $\phi_B(x_1,b_1)$
($\phi_{D^{(*)}}(x_2,b_2)$) has a peak at the momentum fraction
$x_1\equiv k_1^-/P_1^-\sim \bar\Lambda/m_B$ ($x_2\equiv
k_2^+/P_2^+\sim \bar\Lambda/m_{D^{(*)}}$) \cite{TLS2}.

The form factors are then expressed as the convolution of the hard
kernels $H$ with the $B$ and $D^{(*)}$ meson wave functions in
$k_T$ factorization theorem,
\begin{eqnarray}
F^{BD^{(*)}}(q^2)=\int dx_1dx_2 d^2b_1 d^2b_2
\phi_B(x_1,b_1)H(x_1,x_2,b_1,b_2)\phi_{D^{(*)}}(x_2,b_2)\;.
\label{bds}
\end{eqnarray}
The $D^{(*)}$ meson wave function contains a Sudakov factor
arising from $k_T$ resummation, which sums the large double
logarithms $\alpha_s\ln^2(m_Bb_2)$ to all orders. The $B$ meson
wave function also contains such a Sudakov factor, whose effect is
negligible because a $B$ meson is dominated by soft dynamics. The
hard kernels involve a Sudakov factor from threshold resummation,
which sums the large double logarithm $\alpha_s\ln^2 x_1$ or
$\alpha_s\ln^2x_2$ to all orders. This factor modifies the
end-point behavior of the $B$ and $D^{(*)}$ meson wave functions
effectively, rendering them diminish faster in the small $x_{1,2}$
region.

\subsection{End-point Singularity and Sudakov Factor}

It has been pointed out that if evaluating Fig.~\ref{gluon} in 
collinear factorization theorem, an end-point singularity appears 
\cite{SHB}. In this theorem we have the lowest-order hard kernel,
\begin{eqnarray}
H^{(0)}(x_1,x_2)\propto \frac{1}{x_1x_2^2}\;,
\label{hco}
\end{eqnarray}
from the left diagram in Fig.~\ref{gluon}, which leads to a
logarithmic divergence, as the $D^{(*)}$ meson distribution
amplitude behaves like $\phi_{D^{(*)}}(x_2)\propto x_2$ in the small
$x_2$ region. This singularity implies the breakdown of collinear
factorization, and $k_T$ factorization becomes more appropriate. Once
the parton transverse momenta $k_T$ are taken into account,
Eq.~(\ref{hco}) is modified into
\begin{eqnarray}
H^{(0)}(x_1,x_2,{\bf k}_{1T},{\bf k}_{2T})\propto
\frac{m_B^4}{[x_1x_2m_B^2+({\bf k}_{1T}-{\bf k}_{2T})^2]
[x_2m_B^2+{\bf k}_{2T}^2]}\;. \label{hkt}
\end{eqnarray}
A dynamical effect, the so-called Sudakov suppression, favors the
configuration in which $k_T$ is not small \cite{LS}. The end-point
singularity then disappears as explained below.

%%%%%%%%%%%%%%%%%%%%%%%%%%%%%%%%%%%%%%%%%
\begin{figure}[t]
\begin{center}
\epsfig{file=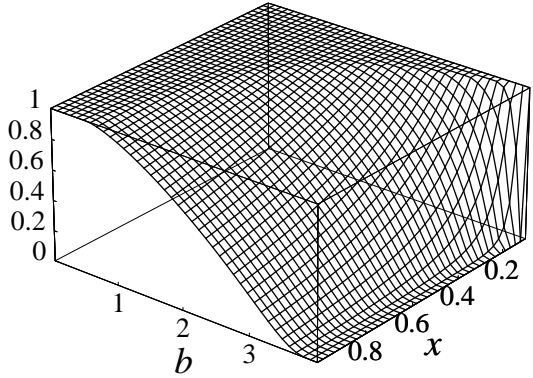,width=7cm}
\end{center}
\caption{QCD demands the presence of a Sudakov factor, which is
an amplitude for a quark-antiquark color dipole not to emit real
gluons in the final state. For a large transverse separation $b$, the
quark and the antiquark do not shield each other's color charge, and 
intend to radiate. In this region Sudakov suppression is strong.}
\label{sudakov}
\end{figure}
%%%%%%%%%%%%%%%%%%%%%%%%%%%%%%%%%%%%%%%%%%%%

When an electron undergoes harder scattering, it intends to
radiate more photons. Hence, the scattering amplitude for
radiating no photons must be suppressed by a factor, whose effect
increases with the electron energy. In QED it is the well-known
Sudakov suppression factor, an amplitude for an electron not to
emit a photon in hard scattering. In the current QCD case of the
$B\to D^{(*)}$ transition, it is the $c$-$\bar d$ quark-antiquark
color dipole that undergoes hard scattering. When
the color dipole is larger, in intends to radiate more gluons.
Since the final state contains only a single $D^{(*)}$ meson, the
real gluon emission is forbidden in the hard decay process.
Similarly, the transition amplitude must involve a Sudakov factor,
whose effect increases with the size of the color dipole, i.e.,
with the separation $b$ between $c$ and $\bar d$. That is, the
configuration with a smaller separation $b$ or with a larger
relative transverse momentum $k_T$ is preferred in the $B\to
D^{(*)}$ transition at large recoil. Then the virtual particles
involved in the hard kernel remain sufficiently off-shell, and
Eq.~(\ref{bds}), with Eq.~(\ref{hkt}) inserted, is free from the
end-point singularity.

The corresponding Sudakov factor can be derived in PQCD as a
function of the transverse separation $b$ and of the momentum
fraction $x$ carried by the spectator quark \cite{LY1}, whose
behavior is shown in Fig.~\ref{sudakov}. The Sudakov factor
suppresses the large $b$ region, where the quark and the antiquark 
are separated by a large transverse distance and the color 
shielding is not effective. It also suppresses the $x\sim 1$ region,
where a quark carries all of the meson momentum, and intends to
emit real gluons in hard scattering. The Sudakov factors from $k_T$
resummation \cite{CS} for the $B$ and $D^{(*)}$ mesons are only
associated with the light spectator quarks, since the double logarithms
arise from the overlap of the soft and mass (collinear) divergences.
These factors, being universal, are the same as in all our previous
analyses.

Similarly, the small $x$ region corresponds to a configuration with
a soft spectator, i.e., with a large color dipole in the longitudinal
direction. The probability for this large color dipole not to radiate
in hard scattering is also described by a Sudakov factor, which
comes from threshold resummation for the hard kernels. For the derivation
of this Sudakov factor, refer to \cite{L5}. For convenience, it has been
parametrized as \cite{TLS},
\begin{eqnarray}
S_t(x)=\frac{2^{1+2c}\Gamma(3/2+c)}{\sqrt{\pi}\Gamma(1+c)}
[x(1-x)]^c\;, \label{ths}
\end{eqnarray}
with the constant $c=0.35$. The above parametrization is motivated by
the qualitative behavior of $S_t$: $S_t(x)\to 0$ as $x\to 0$, 1
\cite{L5}. Since threshold resummation is associated with
the hard kernels, the result could be process-dependent. It has been
observed \cite{LU02} that its effect is essential for factorizable
decay topologies, and negligible for nonfactorizable decay topologies.

\subsection{Factorization Assumption}

We review the basics of FA for the $B\to D^{(*)}M$ decays.
The relevant effective weak Hamiltonian is given by
\begin{eqnarray}
{\cal H}_{\rm eff} = {G_F\over\sqrt{2}}\, V_{cb}V_{ud}^*
\Big[C_1(\mu)O_1(\mu)+C_2(\mu)O_2(\mu)\Big]\;,
\end{eqnarray}
where the four-fermion operators are
\begin{eqnarray}
O_1= (\bar db)_{V-A}(\bar cu)_{V-A}\;,\qquad\qquad
O_2= (\bar cb)_{V-A}(\bar du)_{V-A}\;,
\end{eqnarray}
with the definition $(\bar q_1q_2)_{V-A}\equiv
\bar q_1\gamma_\mu(1- \gamma_5)q_2$, $V$'s the 
Cabibbo-Kobayashi-Maskawa (CKM) matrix
elements, and $C_1$ and $C_2$ the Wilson coefficients.
The $\bar B^0\to D^+\pi^-$ mode is referred to as the class-1
(color-allowed) topology, in which the charged pion is emitted at the
weak vertex. The $\bar B^0\to D^0\pi^0$ mode is referred to as the
class-2 (color-suppressed) topology, in which the $D^0$ meson is
directly produced.

For the $\bar B^0\to D^+\pi^-$ mode, $O_2$ and Fierz transformed
$O_1$ contribute. For the $\bar B^0\to D^0\pi^0$ mode, $O_1$ and
Fierz transformed $O_2$ contribute. Applying FA \cite{BSW} or
generalized FA \cite{Cheng94,Cheng96,Soares}\footnote{The main difference
between FA and generalized FA is that nonfactorizable contributions are
included in the latter.} to the hadronic matrix
elements, we have
\begin{eqnarray}
& &\langle D^+\pi^-|(\bar cb)_{V-A}(\bar du)_{V-A}|\bar B^0\rangle
\approx \langle D^+|(\bar cb)_{V-A}|\bar B^0\rangle
\langle \pi^-|(\bar du)_{V-A}|0\rangle\;,
\nonumber\\
& &\langle D^0\pi^0|(\bar db)_{V-A}(\bar cu)_{V-A}|\bar B^0\rangle
\approx \langle \pi^0|(\bar db)_{V-A}|\bar B^0\rangle
\langle D^0|(\bar cu)_{V-A}|0\rangle\;.
\end{eqnarray}
Substituting the definition of the $B$ meson transition form factors,
$F^{B D}$ and $F^{B\pi}$, and of the meson decay constants, $f_\pi$
and $f_D$, the $\bar B^0\to D^+\pi^-$ (class-1) and 
$\bar B^0\to D^0\pi^0$ (class-2) decay amplitudes are expressed as
\begin{eqnarray}
A(\bar B^0\to D^+\pi^-) &=& i\,\frac{G_F}{\sqrt 2}\,
    V_{cb} V_{ud}^*\,(m_B^2-m_D^2)\,f_\pi\,F^{B D}(m_\pi^2)\,
    a_1(D\pi)\;,
\label{bdf1}\\
\sqrt 2\,A(\bar B^0\to D^0\pi^0) &=& -i\,\frac{G_F}{\sqrt 2}\,
    V_{cb} V_{ud}^*\,(m_B^2-m_\pi^2)\,f_D\,F^{B\pi}(m_D^2)\,
    a_2(D\pi)\;,
\label{bdfa}
\end{eqnarray}
where the parameters $a_1$ and $a_2$ are defined by
\begin{eqnarray}
a_1 = C_2(\mu) + {C_1(\mu)\over N_c}\;,\;\;\;\;
%+ \chi_2(\mu)\right]\;,
%\nonumber\\
a_2& =& C_1(\mu) + {C_2(\mu) \over N_c}\;,
%+\chi_1(\mu)\right]\;,
\label{bsw}
\end{eqnarray}
$N_c$ being the number of colors.
The $B^-\to D^0\pi^-$ mode, involving both classes of amplitudes, is
referred to as class-3. The isospin symmetry implies
\begin{equation}
  A(\bar B^0\to D^+\pi^-)  = A(B^-\to D^0\pi^-)
   + \sqrt 2\,A(\bar B^0\to D^0\pi^0) \;.
\label{iso}
\end{equation}

%In the generalized FA, nonfactorizable contributions are parameterized
%through the phenomenological coefficients $\chi_1$ and $\chi_2$.
%The parameters $a_1$ and $a_2$ in Eq.~(\ref{bdfa}) are then written as
%The $\mu$ dependence of the Wilson coefficients is assumed to be exactly
%compensated by that of $\chi_i(\mu)$ \cite{NRSX}.

It is straightforward to apply FA to other $\bar B\to D^{(*)}M$ modes.
$a_1$ and $a_2$ depend on the color and Dirac structures
of the operators, but otherwise are postulated to be universal
\cite{BSW,NRSX,Dean93}. They have the orders of magnitude
$a_1(D\pi)\sim O(1)$ and $a_2(D\pi)\sim O(1/N_c)$.
The consistency of FA can be tested by
comparing $a_1$ and $a_2$ extracted from various decays.
Within errors, the class-1 decays $\bar B^0\to D^{(*)+} M^-$ with
$M=\pi,\rho,a_1,D_s$, and $D_s^*$ are described using a universal value
$|a_1|\approx 1.1\pm 0.1$, whereas the class-2 decays
$\bar B\to\bar K^{(*)} M$ with $M=J/\psi$ and $\psi(2S)$ suggest a nearly
universal value $|a_2|\approx 0.2$--0.3 \cite{a1a2}. The wide range of
$|a_2|$ is due to the uncertainty in the $B\to K^{(*)}$ form factors.
%in QCD-factorization
%on the other hand, in PQCD it comes from the sharpness of the LCDAs of
%charmonium particles, $J/\psi, \psi(2S),..$\cite{charmonium} }.
The class-3 decays $B^-\to D^{(*)0} M^-$ with $M=\pi$ and $\rho$,
which are sensitive to the interference of the two decay topologies,
can be explained by a real and positive ratio $a_2/a_1\approx 0.2$--0.3,
which seemed to agree with the above determination of $|a_1|$ and $|a_2|$.
This is the reason FA was claimed to work well in
explaining two-body charmed $B$ meson decays, before the class-2 modes
$\bar B^0\to D^0M^0$ with $M=\pi,\eta$, and $\omega$ were measured.

The recently observed $\bar B^0\to D^0M^0$ branching ratios
listed in Table~\ref{dpda} \cite{BelleC,CLEOC} revealed interesting QCD
dynamics. The parameter $|a_2|$ directly extracted from these modes
falls into the range of $|a_2(D\pi)|\sim 0.35-0.60$ and
$|a_2(D^*\pi)|\sim 0.25-0.50$ \cite{C02}. To maintain the predictions
for the class-3 decays, there must exist sizeable relative strong phases
between class-1 and class-2 amplitudes \cite{NPe}, which are
$Arg(a_2/a_1)=59^\circ$ for the $D\pi$ modes and $Arg(a_2/a_1)=63^\circ$
for the $D^*\pi$ modes \cite{C02}. These results can be
regarded as a failure of FA: the parameters $a_2$ in
different types of decays, such as $\bar B\to D^{(*)}\pi$ and
$\bar B\to\bar K^{(*)} J/\psi$, differ by almost a factor 2 in magnitude,
implying strong nonuniversal nonfactorizable effects.
It is then crucial to understand this nonuniversality and, especially,
the mechanism responsible for the large relative phases
in a systematic QCD framework.

\section{$B\to D\pi$ IN PQCD}

In this section we take the $B\to D\pi$ decays as an example of 
the PQCD analysis. The intensive study of all other modes
will be performed in the next section.
The $B\to D\pi$ decay rates have the expressions,
\begin{equation}
\Gamma_i=\frac{1}{128\pi}G_F^2|V_{cb}|^2|V_{ud}|^2\frac{m_B^3}{r}
|A_i|^2\;.
\label{br}
\end{equation}
The indices for the classes
$i=1$, 2, and 3, denote the modes ${\bar B}^0\to D^+\pi^-$,
${\bar B}^0\to D^0\pi^0$, and $B^-\to D^0\pi^-$, respectively. The
amplitudes $A_i$ are written as
\begin{eqnarray}
A_1&=&f_\pi \xi_{\rm ext} + f_B\xi_{\rm exc}+{\cal M}_{\rm ext}
+{\cal M}_{\rm exc}\;,
\label{M1}\\
\sqrt{2}A_2&=& -( f_D\xi_{\rm int} - f_B\xi_{\rm exc}
      +{\cal M}_{\rm int} -{\cal M}_{\rm exc} )\;,
\label{M2}\\
A_3&=&f_\pi\xi_{\rm ext} +
f_D\xi_{\rm int}+{\cal M}_{\rm ext}+{\cal M}_{\rm int}\;,
\label{M3}
\end{eqnarray}
with $f_B$ being the $B$ meson decay constant. The
functions $\xi_{\rm ext}$, $\xi_{\rm int}$, and $\xi_{\rm exc}$
denote the factorizable external $W$-emission (color-allowed),
internal $W$-emission (color-suppressed), and
$W$-exchange contributions, which come from Figs.~4(a) and 4(b),
Figs.~5(a) and 5(b), and Figs.~6(a) and 6(b), respectively.
The functions ${\cal M}_{\rm ext}$,
${\cal M}_{\rm int}$, and ${\cal M}_{\rm exc}$ represent the
nonfactorizable external $W$-emission, internal $W$-emission, and
$W$-exchange contributions, which come from Figs.~4(c) and 4(d),
Figs.~5(c) and 5(d), and Figs.~6(c) and 6(d), respectively.
All the topologies, including the factorizable and nonfactorizable
ones, have been taken into account. It is easy to find that
Eqs.~(\ref{M1})-(\ref{M3}) obey the isospin relation in Eq.~(\ref{iso}).
%%%%%%%%%%%%%%%%%%%%%%%%%%%%%%%%%%%%%%%%%%%%%%%%%%%%%%%%%%%%%%%%%
\begin{figure}[htbp]
%\begin{minipage}[c]{2.5cm}
 \scalebox{0.7}{
 {
   \begin{picture}(140,120)(-30,0)
    \ArrowLine(30,60)(0,60)
    \ArrowLine(60,60)(30,60)
    \ArrowLine(90,60)(60,60)
    \ArrowLine(30,20)(90,20)
    \ArrowLine(0,20)(30,20)
    \Gluon(30,20)(30,60){5}{4} \Vertex(30,20){1.5} \Vertex(30,60){1.5}
 %   \Vertex(60,90){1.5}
 %------- weak vertex --------
    \Line(58,62)(62,58)
    \Line(58,58)(62,62)
 %------- weak vertex --------
    \ArrowLine(45,105)(60,65)
    \ArrowLine(60,65)(75,105)
    \put(-20,35){$B$}
    \put(100,35){$ D^{(*)}$}
    \put(58,110){$\pi$}
    \put(45,48){\small{}}
    \put(65,66){\small{}}
    \put(0,65){\small{$\bar{b}$}}
    \put(45,0){(a)}
 \end{picture}
 }}
%\end{minipage}
 \scalebox{0.7}{
 {
   \begin{picture}(140,120)(-30,0)
      \ArrowLine(30,60)(0,60)
      \ArrowLine(60,60)(30,60)
      \ArrowLine(90,60)(60,60)
      \ArrowLine(60,20)(90,20)
      \ArrowLine(0,20)(60,20)
      \Gluon(60,20)(60,60){5}{4} \Vertex(60,20){1.5} \Vertex(60,60){1.5}
%     \Vertex(60,90){1.5}
%------- weak vertex --------
      \Line(28,62)(32,58)
      \Line(28,58)(32,62)
%------- weak vertex --------
      \ArrowLine(15,105)(30,65)
      \ArrowLine(30,65)(45,105)
      \put(-20,35){$B$}
      \put(100,35){$ D^{(*)}$}
      \put(28,110){$\pi$}
      \put(15,48){\small{}}
      \put(35,66){\small{}}
      \put(0,65){\small{$\bar{b}$}}
      \put(35,0){(b)}
 \end{picture}
 }}
 \scalebox{0.7}{
 {
   \begin{picture}(170,110)(-20,0)
      \ArrowLine(47,60)(0,60)
      \ArrowLine(100,60)(53,60)
      \ArrowLine(20,20)(100,20)
      \ArrowLine(0,20)(20,20)
      \ArrowLine(37,80)(47,60)
      \ArrowLine(27,100)(37,80)
      \ArrowLine(53,60)(73,100)
      \Vertex(20,20){1.5} \Vertex(37,80){1.5}
      \GlueArc(150,20)(130,152,180){4}{9}
      \put(-20,35){$B$}
      \put(110,38){$ D^{(*)}$}
      \put(48,105){$\pi$}
      \put(15,48){\small{}}
      \put(55,48){\small{}}
      \put(0,65){\small{$\bar{b}$}}
      \put(45,0){(c)}
%      \put(175,0){(d)}
%      \put(130,50){$+$}
   \end{picture}
   \begin{picture}(140,110)(-20,0)
      \ArrowLine(47,60)(0,60)
      \ArrowLine(100,60)(53,60)
      \ArrowLine(80,20)(100,20)
      \ArrowLine(0,20)(80,20)
      \ArrowLine(27,100)(47,60)
      \ArrowLine(63,80)(73,100)
      \ArrowLine(53,60)(63,80)
      \Vertex(80,20){1.5} \Vertex(63,80){1.5}
      \GlueArc(-50,20)(130,0,28){4}{9}
      \put(-20,35){$B$}
      \put(110,38){$ D^{(*)}$}
      \put(48,105){$\pi$}
      \put(15,48){\small{}}
      \put(55,48){\small{}}
      \put(0,65){\small{$\bar{b}$}}
      \put(45,0){(d)}
   \end{picture}
 }}
\vskip 0.5cm
 \caption{Color-allowed emission diagrams contributing to the
 $B\to D^{(*)}\pi$ decays.}
\end{figure}
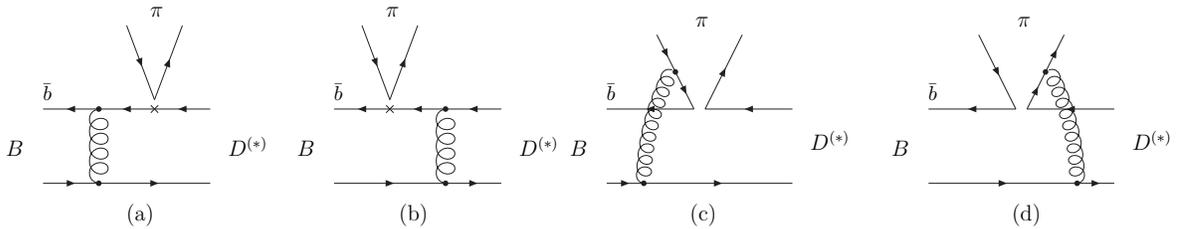

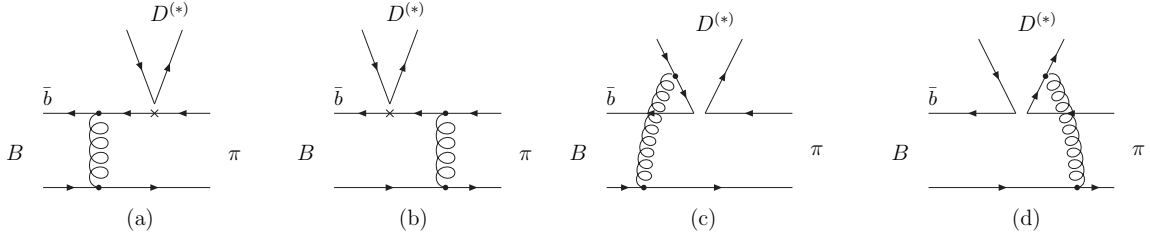
\begin{figure}[htbp]
%\begin{minipage}[c]{2.5cm}
 \scalebox{0.7}{
 {
   \begin{picture}(140,120)(-30,0)
    \ArrowLine(30,60)(0,60)
    \ArrowLine(60,60)(30,60)
    \ArrowLine(90,60)(60,60)
    \ArrowLine(30,20)(90,20)
    \ArrowLine(0,20)(30,20)
    \Gluon(30,20)(30,60){5}{4} \Vertex(30,20){1.5} \Vertex(30,60){1.5}
 %   \Vertex(60,90){1.5}
 %------- weak vertex --------
    \Line(58,62)(62,58)
    \Line(58,58)(62,62)
 %------- weak vertex --------
    \ArrowLine(45,105)(60,65)
    \ArrowLine(60,65)(75,105)
    \put(-20,35){$B$}
    \put(100,35){$\pi$}
    \put(58,110){$ D^{(*)}$}
    \put(45,48){\small{}}
    \put(65,66){\small{}}
    \put(0,65){\small{$\bar{b}$}}
    \put(45,0){(a)}
 \end{picture}
 }}
%\end{minipage}
 \scalebox{0.7}{
 {
   \begin{picture}(140,120)(-30,0)
      \ArrowLine(30,60)(0,60)
      \ArrowLine(60,60)(30,60)
      \ArrowLine(90,60)(60,60)
      \ArrowLine(60,20)(90,20)
      \ArrowLine(0,20)(60,20)
      \Gluon(60,20)(60,60){5}{4} \Vertex(60,20){1.5} \Vertex(60,60){1.5}
%     \Vertex(60,90){1.5}
%------- weak vertex --------
      \Line(28,62)(32,58)
      \Line(28,58)(32,62)
%------- weak vertex --------
      \ArrowLine(15,105)(30,65)
      \ArrowLine(30,65)(45,105)
      \put(-20,35){$B$}
      \put(100,35){$\pi$}
      \put(28,110){$ D^{(*)}$}
      \put(15,48){\small{}}
      \put(35,66){\small{}}
      \put(0,65){\small{$\bar{b}$}}
      \put(35,0){(b)}
 \end{picture}
 }}
 \scalebox{0.7}{
 {
   \begin{picture}(170,110)(-20,0)
      \ArrowLine(47,60)(0,60)
      \ArrowLine(100,60)(53,60)
      \ArrowLine(20,20)(100,20)
      \ArrowLine(0,20)(20,20)
      \ArrowLine(37,80)(47,60)
      \ArrowLine(27,100)(37,80)
      \ArrowLine(53,60)(73,100)
      \Vertex(20,20){1.5} \Vertex(37,80){1.5}
      \GlueArc(150,20)(130,152,180){4}{9}
      \put(-20,35){$B$}
      \put(110,38){$\pi$}
      \put(48,105){$ D^{(*)}$}
      \put(15,48){\small{}}
      \put(55,48){\small{}}
      \put(0,65){\small{$\bar{b}$}}
      \put(45,0){(c)}
%      \put(175,0){(d)}
%      \put(130,50){$+$}
   \end{picture}
   \begin{picture}(140,110)(-20,0)
      \ArrowLine(47,60)(0,60)
      \ArrowLine(100,60)(53,60)
      \ArrowLine(80,20)(100,20)
      \ArrowLine(0,20)(80,20)
      \ArrowLine(27,100)(47,60)
      \ArrowLine(63,80)(73,100)
      \ArrowLine(53,60)(63,80)
      \Vertex(80,20){1.5} \Vertex(63,80){1.5}
      \GlueArc(-50,20)(130,0,28){4}{9}
      \put(-20,35){$B$}
      \put(110,38){$\pi$}
      \put(48,105){$ D^{(*)}$}
      \put(15,48){\small{}}
      \put(55,48){\small{}}
      \put(0,65){\small{$\bar{b}$}}
      \put(45,0){(d)}
   \end{picture}
 }}
\vskip 0.5cm
 \caption{Color-suppressed emission diagrams contributing to the
 $B\to D^{(*)}\pi$ decays.}
\end{figure}

\begin{figure}[htbp]
%\begin{minipage}[c]{2.5cm}
 \scalebox{0.7}{
 {
   \begin{picture}(130,120)(0,-20)
      \put(45,0){
      \rotatebox{90}{
        \ArrowLine(30,60)(0,60)
        \ArrowLine(60,60)(30,60)
        \ArrowLine(90,60)(60,60)
        \ArrowLine(30,20)(90,20)
        \ArrowLine(0,20)(30,20)
        \Gluon(30,20)(30,60){5}{4} \Vertex(30,20){1.5} \Vertex(30,60){1.5}
%------- weak vertex --------
        \Line(58,62)(62,58)
        \Line(58,58)(62,62)
%------- weak vertex --------
        \ArrowLine(45,105)(60,65)
        \ArrowLine(60,65)(75,105)
        \put(65,66){\small{}}
      }}
      \put(0,55){$B$}
      \put(80,95){$ D^{(*)}$}
      \put(80,-10){$\pi$}
      \put(15,78){\small{$\bar{b}$}}
      \put(65,55){\small{}}
%      \put(115,45){$+$}
      \put(60,-30){(a)}
   \end{picture}
   \begin{picture}(110,120)(0,-20)
      \put(85,0){
      \rotatebox{90}{
        \ArrowLine(30,60)(0,60)
        \ArrowLine(60,60)(30,60)
        \ArrowLine(90,60)(60,60)
        \ArrowLine(60,20)(90,20)
        \ArrowLine(0,20)(60,20)
        \Gluon(60,20)(60,60){5}{4} \Vertex(60,20){1.5} \Vertex(60,60){1.5}
%------- weak vertex --------
        \Line(28,62)(32,58)
        \Line(28,58)(32,62)
%------- weak vertex --------
        \ArrowLine(15,105)(30,65)
        \ArrowLine(30,65)(45,105)
        \put(35,66){\small{}}
      }}
      \put(40,25){$B$}
      \put(120,95){$ D^{(*)}$}
      \put(100,-10){$\pi$}
      \put(55,48){\small{$\bar{b}$}}
      \put(105,25){\small{}}
       \put(100,-30){(b)}
         \end{picture}
 }}
 \scalebox{0.7}{
 {
   \begin{picture}(110,130)(0,-20)
      \put(195,0){
      \rotatebox{90}{
        \ArrowLine(47,60)(0,60)
        \ArrowLine(100,60)(53,60)
        \ArrowLine(80,20)(100,20)
        \ArrowLine(0,20)(80,20)
        \ArrowLine(27,100)(47,60)
        \ArrowLine(63,80)(73,100)
        \ArrowLine(53,60)(63,80)
        \Vertex(80,20){1.5} \Vertex(63,80){1.5}
        \GlueArc(-50,20)(130,0,28){4}{9}
      }}
      \put(80,45){$B$}
      \put(155,-10){$\pi$}
      \put(155,105){$ D^{(*)}$}
      \put(140,37){\small{}}
      \put(140,55){\small{}}
      \put(100,75){\small{$\bar{b}$}}
      \put(155,-30){(c)}
         \end{picture}
   \begin{picture}(140,130)(0,-20)
      \put(235,0){
      \rotatebox{90}{
        \ArrowLine(47,60)(0,60)
        \ArrowLine(100,60)(53,60)
        \ArrowLine(20,20)(100,20)
        \ArrowLine(0,20)(20,20)
        \ArrowLine(37,80)(47,60)
        \ArrowLine(27,100)(37,80)
        \ArrowLine(53,60)(73,100)
        \Vertex(20,20){1.5} \Vertex(37,80){1.5}
        \GlueArc(150,20)(130,152,180){4}{9}
      }}
      \put(120,45){$B$}
      \put(195,-10){$\pi$}
      \put(195,105){$ D^{(*)}$}
      \put(180,38){\small{}}
      \put(180,55){\small{}}
      \put(140,75){\small{$\bar{b}$}}
%      \put(115,45){$+$}
      \put(195,-30){(d)}
   \end{picture}
 }}
\vskip 0.5cm
 \caption{Annihilation Diagrams contributing to the $B\to D^{(*)}\pi$
 decays.}
\end{figure}
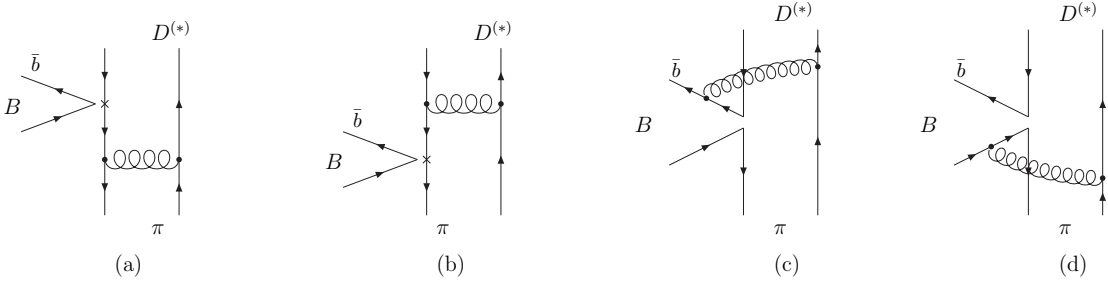
%%%%%%%%%%%%%%%%%%%%%%%%%%%%%%%%%%%%%%%%%%%%%%%%%%%%%%%%%%%%

In the PQCD framework based on $k_T$ factorization theorem, an
amplitude is expressed as the convolution of hard $b$ quark decay
kernels with meson wave functions in both the longitudinal
momentum fractions and the transverse momenta of partons. Our PQCD
formulas are derived up to leading-order in $\alpha_s$, to leading
power in $m_D/m_B$ and in $\bar\Lambda/m_D$, and to leading
double-logarithm resummations. For the Wilson coefficients, we
adopt the leading-order renormalization-group evolution for
consistency, although the next-to-leading-order ones are available
in the literature \cite{buras}. For the similar reason, we employ
the one-loop running coupling constant
$\alpha_s(\mu)=2\pi/[\beta_1 \ln(\mu/{\Lambda_{\rm
QCD}^{(n_f)}})]$ with $\beta_1=(33-2n_f)/3$, $n_f$ being the
number of active quarks. The QCD scale is chosen as $\Lambda_{\rm
QCD}^{(5)}=193$ MeV for the scale $ m_b< \mu< m_W$, which is
derived from $\Lambda_{\rm QCD}^{(4)}=250$ MeV for $\mu < m_b$.

The leading-order and leading-power factorization formulas for the
above decay amplitudes are collected in Appendix A. Here we
mention only some key ingredients in the calculation. The formulas
for the $B\to D\pi$ decays turn out to be simpler than those for
the $B\to\pi\pi$ ones. The simplicity is attributed to the power
counting rules under the hierachy of the three scales in
Eq.~(\ref{hie}). The hard kernels are evaluated up to the power
corrections of order $\bar\Lambda/m_D\sim m_D/m_B$
($\bar\Lambda/m_B$ is regarded as being of even higher power).
Following these rules, the terms proportional to $x_1\sim
\bar\Lambda/m_B$ and to $x_2\sim \bar\Lambda/m_D$ are higher-power
compared to the leading $O(1)$ terms and dropped. We have also
dropped the terms of higher powers in $r=m_D/m_B$. Accordingly,
the phase space factor $1-r^2$, appearing in Eq.~(\ref{br})
originally, has been approximated by 1. This approximation is
irrelevant for explaining the ratios of the $B\to D\pi$ branching
ratios, and causes an uncertainty in the absolute branching
ratios, which is much smaller than those from the CKM matrix
element $|V_{cb}|$, and from the meson decay constants $f_B$ and
$f_D$.

Up to the power corrections of order $\bar\Lambda/m_B$ and
$\bar\Lambda/m_D$, we consider only a single $B$ ($D$) meson
wave function. The nonperturbative $B$ meson and pion wave functions
have been fixed in our previous works \cite{KLS,LUY}. The unknown $D$
meson wave function was determined by fitting the PQCD predictions for
the $B\to D$ transition form factors to the observed $B\to Dl\nu$ decay
spectrum \cite{TLS2}. The contributions from the two-parton twist-3
$D$ meson wave functions, being higher-power, are
negligible. Note that in the charmless decays the contributions
from the two-parton twist-3 light meson distribution amplitudes are
not down by a power of $1/m_B$. These
distribution amplitudes, being constant at the momentum fraction $x\to 0$
as required by equations of motion, lead to linear singularities in
the collinear factorization formulas. The linear singularities modify
the naive power
counting, such that two-parton twist-3 contributions become leading-power
\cite{TLS,CKL}. In the charmed decays the above equations of motion are
modified \cite{TLS2}, and the two-parton twist-3 $D$ meson wave
functions vanish at the end point of $x$. Therefore, their contributions
are indeed higher-power.

Retaining the parton transverse momenta $k_T$, the nonfactorizable
topologies generate strong phases from non-pinched singularities
of the hard kernels \cite{WYL}. For example, the virtual quark
propagator in Fig.~5(d) is written, in the principle-value
prescription, as
\begin{eqnarray}
\frac{1}{x_3(x_2-x_1)m_B^2-({\bf k}_{2T}-{\bf k}_{1T}+{\bf k}_{3T})^2
+i\epsilon}
&=&P\left[\frac{1}{x_3(x_2-x_1)m_B^2-({\bf k}_{2T}-{\bf k}_{1T}+{\bf
k}_{3T})^2}\right]
\nonumber\\
& &-i\pi\delta(x_3(x_2-x_1)m_B^2-({\bf k}_{2T}-{\bf k}_{1T}
+{\bf k}_{3T})^2)\;,
\nonumber\\
& &
\label{pri}
\end{eqnarray}
with $x_3$ being the momentum fraction associated with the
pion. The second term contributes to the strong phase, which is thus
of short-distance origin and calculable.  The first
term in the above expression does not lead to an end-point singularity.
Note that the strong phase from Eq.~(\ref{pri}) is obtained by keeping
all terms in the denominator of a propagator without neglecting $x_1$
and $x_2$.

\section{NUMERICAL RESULTS}

The computation of the hard kernels in $k_T$ factorization theorem
for other charmed decay modes is similar and straightforward. The
$B\to D^*\pi$ decay amplitudes are the same as the $B\to D\pi$
ones but with the substitution of the mass, the decay constant,
and the distribution amplitude,
\begin{eqnarray}
m_D\to m_{D^*}\;,\;\;f_{D}\to f_{D^*}\;,\;\; \phi_{D}(x_2)\to
\phi_{D^*}(x_2)\;.
\label{rep1}
\end{eqnarray}
This simple substitution is expected at leading power under the
hierachy in  Eq.~(\ref{hie}): the difference between the two
channels should occur only at $O(\bar\Lambda/m_D)$. An explicit
derivation shows that the difference occurs at the twist-3 level
for the nonfactorizable emission diagrams in Fig.~4 and for the
annihilation diagrams in Fig.~6. It implies that the universality
(channel-independence) of $a_1$ and $a_2$ assumed in FA \cite{fac}
breaks down at subleading power even within the $B\to D^{(*)}M$
decays.

Replacing the pion in Figs.~4-6 by the $\rho$ ($\omega$) meson, we
obtain the diagrams for the $B\to D^{(*)}\rho (\omega)$ decays.
The factorization formulas for the $B\to D\rho(\omega)$ decay
amplitudes are also the same as those for the $B\to D\pi$ ones but
with the substitution,
\begin{eqnarray}
\phi_\pi\to\phi_{\rho,\omega}\;,\;\;\;\;
\phi_\pi^p\to\phi_{\rho,\omega}^s\;,\;\;\;\;
\phi_\pi^t\to\phi_{\rho,\omega}^t\;,\;\;\;\; m_0\to
m_{\rho(\omega)}\;,
\end{eqnarray}
where $\phi_\pi$ ($\phi_\pi^{p,t}$) is the two-parton twist-2
(twist-3) pion distribution amplitude, $\phi_{\rho,\omega}$
($\phi_{\rho,\omega}^{s,t}$) the two-parton twist-2 (twist-3)
$\rho$ and $\omega$ meson distribution amplitudes, respectively,
$m_0$ the chiral
enhancement scale, and $m_{\rho(\omega)}$ the $\rho$ ($\omega$)
meson mass. Hence, the similar isospin relation holds:
\begin{equation}
A(B^0 \to D^{-}\rho^+)=A(B^+ \to\bar D^{0}\rho^+)
+\sqrt{2}A(B^0 \to\bar D^{0}\rho^0)\;.
\end{equation}

The $B\to D^{*}\rho (\omega)$ decays contain more amplitudes
associated with the different polarizations.
%, whose explicit expressions are referred to Appendix B.
However, at leading power, the amplitudes associated with the
transverse polarizations, suppressed by a power of $m_{D^*}/m_B$ 
or of $m_\rho/m_B$, are negligible. That is, the factorization 
formulas for the $B\to D^{*}\rho (\omega)$ modes are the same as the
$B\to D\rho (\omega)$ ones with the substitution in Eq.~(\ref{rep1}).
Note that the $W$-exchange contribution changes sign in the 
$\bar B^0\to D^{(*)0}\omega$ decay amplitude:
\begin{eqnarray}
A(\bar B^0\to D^{(*)0}\omega) =
-\frac{1}{\sqrt{2}}(f_{D^{(*)}}\xi_{\rm int} + f_B\xi_{\rm exc} +
{\cal M}_{\rm int} + {\cal M}_{\rm exc})\;, \label{dom}
\end{eqnarray}
due to the different quark structures between
the $\omega$ meson (proportional to $u \bar u + d \bar d$) and
the $\rho^0$ meson (proportional to $u \bar u - d \bar d$).

In the numerical analysis we adopt the model for the $B$ meson
wave function,
\begin{eqnarray}
\phi_B(x,b)=N_Bx^2(1-x)^2
\exp\left[-\frac{1}{2}\left(\frac{xm_B}{\omega_B}\right)^2
-\frac{\omega_B^2 b^2}{2}\right]\;,
\label{os}
\end{eqnarray}
with the shape parameter $\omega_B$ and the normalization constant
$N_B$ being related to the decay constant $f_B$ through
\begin{eqnarray}
\int dx\phi_B(x,0)=\frac{f_B}{2\sqrt{2N_c}}\;.
\end{eqnarray}
The $D^{(*)}$ meson distribution amplitude is given by
\begin{eqnarray}
\phi_{D^{(*)}}(x)=\frac{3}{\sqrt{2N_c}}f_{D^{(*)}}
x(1-x)[1+C_{D^{(*)}}(1-2x)]\;,
%\exp[-(\omega_D b)^2/2]
\label{phid}
\end{eqnarray}
with the shape parameter $C_{D^{(*)}}$. The pion and $\rho(\omega)$ 
meson distribution amplitudes have been derived in \cite{PB1,PB2}, 
whose explicit expressions are given in Appendix A. The $B$ meson 
wave function was then extracted from the light-cone-sum-rule (LCSR) 
results of the $B\to\pi$ transition form factor \cite{TLS}. The 
range of $C_{D^{(*)}}$ was determined from the measured 
$B\to D^{(*)} l\nu$ decay spectrum at large recoil employing the 
$B$ meson wave function extracted above. We do not consider the 
variation of $\phi_{D^{(*)}}$ with the impact parameter $b$, since 
the available data are not yet sufficiently precise to control
this dependence.

The input parameters are listed below:
\begin{eqnarray}
& & f_B = 190\; {\rm MeV}\;,\;\; \omega_B=0.4\;{\rm GeV}\;,
\nonumber\\
& &f_{D} = 240\; {\rm MeV}\;,\;\;C_D=0.8\pm 0.2\;,
\nonumber\\
& &f_{D^*} = 230\; {\rm MeV}\;,C_{D^*}=0.7\pm 0.2\;,
\nonumber\\
& & f_\pi = 132\;{\rm MeV}\;,\;\;f_\rho = f_\omega=200\;{\rm MeV}\;,\;\;
 f_\rho^T =f_\omega^T = 160\;{\rm MeV}\;,
\nonumber\\
& &m_B = 5.28\; {\rm GeV}\;,\;\; m_b = 4.8\; {\rm GeV}\;,\;\;
\nonumber\\
& &m_D = 1.87\; {\rm GeV}\;,\;\; m_{D^*} = 2.01\; {\rm GeV}\;,\;\;
m_c = 1.3\; {\rm GeV}\;,\;\;
\nonumber\\
& &m_\rho=0.77\; {\rm GeV}\;,\;\; m_\omega=0.78\;{\rm GeV}\;,
\nonumber\\
& &m_t=170\; {\rm GeV}\;,\;\;m_W = 80.41\;{\rm GeV}\;,\;\; m_0 =
1.4\; {\rm GeV}\;,
\nonumber\\
& &\tau_{B^\pm}=1.674\times 10^{-12}{\rm s}\;,\;\;
\tau_{B^0}=1.542\times 10^{-12}{\rm s}\;,
\nonumber\\
& &G_F=1.16639\times 10^{-5}\;{\rm GeV}^{-2}\;,\;\;
|V_{cb}|=0.043\;,\;\;|V_{ud}|=0.974\;, \label{para}
\end{eqnarray}
where $m_t$, $m_W$, $\tau_{B^\pm}$ and $\tau_{B^0}$ denote the top
quark mass, the $W$ boson mass, the $B^\pm$ meson lifetime, and
the $B^0$ meson lifetime, respectively. The above meson wave
functions and parameters correspond to the form factors at maximal
recoil,
\begin{eqnarray}
F_0^{B\pi}\sim 0.3\;,~~~~~~ \xi_+^{BD} \sim 0.57\;,~~~~~~
\xi_{A_1}^{BD^{*}} \sim 0.52\;,\label{ffl}
\end{eqnarray}
which are close to
the results from QCD sum rules \cite{sum,BG}. We
stress that there is no arbitrary parameter in our calculation,
though the value of each parameter is only known up to a range.

The PQCD predictions for each term of the $B\to D\pi$ decay
amplitudes are exhibited in Table~\ref{dpt}. The theoretical
uncertainty comes only from the variation of the shape parameter
for the $D$ meson distribution amplitude, $0.6 < C_D < 1.0$. It is
expected that the color-allowed factorizable amplitude
$f_\pi\xi_{\rm ext}$ dominates, and that the color-suppressed
factorizable contribution $f_D\xi_{\rm int}$ is smaller due to the
Wilson coefficient $C_1+C_2/N_c\sim 0$. The color-allowed
nonfactorizable amplitude ${\cal M}_{\rm ext}$ is negligible:
since the pion distribution amplitude is symmetric under the
exchange of $x_3$ and $1-x_3$, the contributions from the two
diagrams Figs.~4(c) and 4(d) cancel each other in the dominant
region with small $x_2$. It is also down by the small Wilson
coefficient $C_1/N_c$. For the color-suppressed nonfactorizable
contribution ${\cal M}_{\rm int}$, the above cancellation does not
exist in the dominant region with small $x_3$, because the $D$
meson distribution amplitude $\phi_D(x_2)$ is not symmetric under
the exchange of $x_2$ and $1-x_2$. Furthermore, ${\cal M}_{\rm
int}$, proportional to $C_2/N_c\sim 0.3$, is not down by the
Wilson coefficient. It is indeed comparable to the color-allowed
factorizable amplitude $f_\pi\xi_{\rm ext}$, and produces a large
strong phase as explained in Eq.~(\ref{pri}). Both the
factorizable and nonfactorizable annihilation contributions are
small, consistent with our argument in Sec.~II.

The predicted branching ratios in Table~\ref{dp} are in agreement
with the averaged experimental data \cite{BelleC,CLEOC,Bab}. We
extract the parameters $a_1$ and $a_2$ by equating
Eqs.~(\ref{bdf1}) and (\ref{bdfa}) to Eqs.~(\ref{M1}) and
(\ref{M2}), respectively. That is, our $a_1$ and $a_2$ do not only
contain the nonfactorizable amplitudes as in generalized FA, but
the small annihilation amplitudes, which was first discussed in
\cite{GKKP}. We obtain the ratio $|a_2/a_1|\sim 0.43$ with $10\%$
uncertainty and the phase of $a_2$ relative to $a_1$ about
$Arg(a_2/a_1)\sim -42^\circ$. If excluding the annihilation
amplitudes $f_B\xi_{\rm exc}$ and ${\cal M}_{\rm exc}$, we have
$|a_2/a_1|\sim 0.46$ and $Arg(a_2/a_1)\sim -64^\circ$. Note that
the experimental data do not fix the sign of the relative phases.
The PQCD calculation indicates that $Arg(a_2/a_1)$ should be
located in the fourth quadrant. It is evident that the
short-distance strong phase from the color-suppressed
nonfactorizable amplitude is already sufficient to account for the
isospin triangle formed by the $B\to D\pi$ modes. The conclusion
that the data hint large final-state interaction was drawn from
the analysis based on FA \cite{NPe,C02,X,CHY}. Hence, it is more
reasonable to claim that the data just imply a large strong phase,
but do not tell what mechanism generates this phase \cite{JPL}.
From the viewpoint of PQCD, this strong phase is of short
distance, and produced from the non-pinched singularity of the
hard kernel. Certainly, under the current experimental and
theoretical uncertainties, there is still room for long-distance
phases from final-state interaction.

The PQCD predictions for the $B\to D^*\pi$ decay amplitudes and
branching ratios in Table~\ref{dsp} are also consistent with the
data \cite{PDG}. Since $m_{D^*}$ and $\phi_{D^*}$ are only slightly
different from $m_{D}$ and $\phi_{D}$, respectively, the results
are close to those in Table~\ref{dp}. The $B\to D^*\pi$ branching
ratios are smaller than the $B\to D\pi$ ones because of the form
factors $\xi^{BD^*}_{A_1}<\xi^{BD}_+$ as shown in Eq.~(\ref{ffl}).
Similarly, the ratio $|a_2/a_1|$ and the relative phase
$Arg(a_2/a_1)$ are also close to those associated with the $B\to
D\pi$ decays. We obtain the ratio $|a_2/a_1|\sim 0.47$ with $10\%$
uncertainty and the relative phase about $Arg(a_2/a_1)\sim
-41^\circ$. Excluding the annihilation amplitudes, we have
$|a_2/a_1|\sim 0.5$ and $Arg(a_2/a_1)\sim -63^\circ$. 

The PQCD
predictions for the $B\to D^{(*)}\rho(\omega)$ branching ratios
are listed in Table~\ref{dr}, which match the data
\cite{PDG}. The $\bar B^0\to D^+\rho^-$ and $B^-\to D^0\rho^-$
branching ratios are about twice of the $\bar B^0\to D^+\pi^-$ 
and $B^-\to D^0\pi^-$ ones because of the larger $\rho$ meson 
decay constant, $(f_\rho/f_\pi)^2\sim 2$. The relatively smaller 
$\bar B^0\to D^{0}\rho^0$ branching ratio is attributed to
the cancellation of the above enhancing effect between the 
color-suppressed and $W$-exchange contributions, consistent with 
the observation made in an analysis based on
the topological-amplitude parametrization \cite{CR}. The $\bar
B^0\to D^{0}\omega$ branching ratio is larger than the $\bar
B^0\to D^{0}\rho^0$ one due to the constructive
inteference between the color-suppressed contribution and the
annihilation contribution as indicated in Eq.~(\ref{dom}). 
To obtain the ${\bar B}^0\to D^{*}\rho$ helicity amplitudes and 
their relative phases \cite{SSU}, the power-suppressed contributions 
from the transverse polarizations must be included. For consistency,
the contribution from the longitudinal polarization should be
calculated up to the same power. We shall study this subject in
a forthcoming paper. The predictions for the $B\to D^{*0}\rho^0$, 
$D^{*0}\omega$ decays can be compared with future measurement. 
The latter branching ratio is larger than the former one because 
of the same reason as for the $B\to D^{0}\rho^0$, 
$D^{0}\omega$ decays.

\section{COMPARISON WITH OTHER APPROACHES}

In this section we make a brief comparison of the PQCD formalism
with other QCD approaches to exclusive $B$ meson decays,
emphasizing the differences. For more details, refer to
\cite{Li03}. As mentioned before, there are two kinds of
factorization theorem for QCD processes \cite{NL}: collinear
factorization, on which the soft-collinear effective theory (SCET)
\cite{bfl,bfps}, LCSR \cite{CZ90,ABS}, and
QCDF \cite{BBNS} are based, and $k_T$ factorization, on which
the PQCD approach is based. Calculating the $B\to\pi$ form
factor $F^{B\pi}$ in collinear factorization up to leading power
in $1/m_B$ and leading-order in $\alpha_s$, an end-point
singularity occurs. Hence, we define three types of contributions:
a genuine soft contribution $f^{\rm S}$, a contribution $f^{\rm
EP}$ containing the end-point singularity, and a finite contribution
$f^{\rm F}$:
\begin{eqnarray}
F^{B\pi}=f^{\rm S}+f^{\rm EP}+f^{\rm F}\;.
\end{eqnarray}
The second term can not cover the complete soft contribution, because 
it is from a leading formalism. Note that the end-point
singularity exists even in the heavy quark limit. Hence, $B$ meson
decays differ from other exclusive processes, which become
calculable in collinear factorization at sufficiently large
momentum transfer.

There are two options to handle the above end-point singularity
\cite{Lmon}: first, an end-point singularity in collinear
factorization implies that exclusive $B$ meson decays are
dominated by soft dynamics. Therefore, a heavy-to-light form
factor is not calculable, and $f^{\rm EP}$ should be treated as a
soft object, like $f^{\rm S}$. In SCET and QCDF,
$F^{B\pi}$ is then written, up to $O(\alpha_s)$, as \cite{PS02,BF}
\begin{eqnarray}
F^{B\pi}= f^{\rm NF}+f^{\rm F} \;, \label{soh}
\end{eqnarray}
with
\begin{eqnarray}
f^{\rm NF}&=&f^{\rm S}+f^{\rm EP}\;, \nonumber\\
f^{\rm F}&=&\phi_B \otimes T' \otimes \phi_{\pi}\;. \label{fff}
\end{eqnarray}
The soft form factor $f^{\rm NF}$, obeying the large-energy
symmetry relations \cite{CYO}, can be estimated in terms of a
triangle diagram without a hard gluon exchange in LCSR 
\cite{sum, Ball03}.
However, since the pion vertex has been replaced by the pion
distribution amplitudes under twist expansion, what is calculated
in LCSR is not the full soft contribution. The term $f^{\rm
F}$ has been expressed as a convolution of the hard-scattering
kernel $T'$ with the light-cone distribution amplitudes of the $B$
meson and of the pion in the momentum fractions, implying that 
it is calculable in collinear factorization.

The second option is that an end-point singularity indicates the 
breakdown of the collinear factorization. Hence, 
the $k_T$ factorization 
is the more appropriate framework, in which the parton
transverse momenta $k_T$ are retained in the hard kernel, and
$f^{\rm EP}$ does not develop an end-point singularity. 
Both $f^{\rm EP}$ and $f^{\rm F}$ are then calculable, and expressed,
in the PQCD approach, as
\begin{eqnarray}
F^{B\pi}= f^{\rm EP}+f^{\rm F}= \phi_B \otimes T \otimes
\phi_{\pi}\;, \label{hd}
\end{eqnarray}
where the symbol $\otimes$ represents the convolution not only in
the momentum fractions, but in the transverse separations. 
The hard kernel $T'$ in Eq.~(\ref{fff}) is derived from the
complete hard kernel $T$ by dropping the terms which lead to the
end-point singularity in the collinear factorization. Certainly, the 
subtraction of these terms depends on a regularization scheme 
\cite{BF}. The strong Sudakov suppression in the soft parton region 
implies that the genuine soft contribution $f^{\rm S}$ is not 
important \cite{LS,TLS}. Equation (\ref{hd}) is then claimed to be a
consequence of the hard-dominance picture, because a big portion
of $F^{B\pi}$ is calculable. The agreement between the sum-rule and
PQCD predictions for many $B$ meson transition form factors 
justifies that $f^{\rm S}$ is indeed
negligible. Since $f^{\rm EP}$ remains in Eq.~(\ref{hd}), the form
factor symmetry relations at large recoil are still respected in
the PQCD framework \cite{TLS}, which are then modified by the 
subleading term $f^{\rm F}$.

Therefore, the soft-dominance (hard-dominance) picture
postulated in LCSR (PQCD) makes sense in the collinear ($k_T$)
factorization \cite{Li03}. The two pictures arise from the 
different theoretical frameworks, and there is no conflict at all. 
In other words, the soft contribution
refers to $f^{\rm NF}$ in SCET, LCSR, and QCDF, which is large,
but to $f^{\rm S}$ in PQCD, which is small. LCSR can be
regarded as a method to evaluate $f^{\rm EP}$ or $f^{\rm NF}$ in
the collinear factorization (at least the light-cone
distribution amplitudes have been employed on the pion side),
while the $k_T$ factorization is adopted in PQCD for the evaluation
of $f^{\rm EP}$. We emphasize that there is no preference
between the two options for semileptonic $B$ meson decays, both of
which give similar results as stated above. However, in their
extension to two-body nonleptonic $B$ meson decays, predictions
could be very different. For example, the main source of strong
phases in the $B\to\pi\pi$ decays is the correction to the weak
vertex in QCDF, but the annihilation diagram in PQCD. This is the
reason QCDF (PQCD) predicts a smaller and positive (larger and
negative) CP asymmetry $C_{\pi\pi}$ \cite{NL,L365}. It is then
possible to discriminate experimentally which theoretical
framework works better.

Next we compare our formalism for two-body charmed nonleptonic $B$
meson decays based on the $k_T$ factorization with SCET and QCDF 
based on the collinear factorization. Currently, LCSR has not yet 
been applied to the nonfactorizable contribution discussed here, 
but only to that from three-parton distribution amplitudes
\cite{Khodja}, since the former, involving two loops, is more
complicated to analyze. Similarly, the neglect of $k_T$ results in
end-point singularities in the factorizable contributions
$\xi_{\rm ext}$ and $\xi_{\rm int}$, which need to be parametrized
in terms of the $B\to D$ and $B\to\pi$ transition form factors,
respectively. It also causes an end-point singularity in the
color-suppressed nonfactorizable amplitude ${\cal M}_{\rm int}$,
if the $c$ quark is treated as being massive. This is why the
color-suppressed modes, i.e., the magnitude and the phase of $a_2$,
can not be predicted in QCDF, and the proof of QCDF in the SCET
formalism \cite{bps} considered only the color-allowed mode $\bar
B^0\to D^+\pi^-$. The color-allowed nonfactorizable amplitude
${\cal M}_{\rm ext}$ is calculable in QCDF, because the end-point
singularities cancel between the pair of diagrams, Figs.~4(c) and
4(d). We mention a recent work on SCET \cite{BPS03}, in which the
color-suppressed nonfactorizable amplitude has been parametrized
as an expression similar to Eq.~(\ref{soh}).

If the $c$ quark is treated as being massless, the end-point
singularities in the pair of color-suppressed nonfactorizable
diagrams, Figs.~5(c) and 5(d), will cancel each other as in the
charmless case \cite{CKL,L365}. This can be understood by
examining the behavior of the integrand of ${\cal M}_{\rm int}$ in
Eq.~(\ref{md}) in the dominant region with small $x_3$, noticing
that the $D$ meson distribution amplitude $\phi_D(x_2)$ would be
symmetric under the exchange of $x_2$ and $1-x_2$ in the $m_c\to
0$ limit. However, the nonfactorizable contribution will become
negligible in this limit, such that the amplitude ${\cal M}_{\rm
int}$, though calculable in QCDF, is not large enough to explain
the $B\to D\pi$ data. It is then obvious that the PQCD approach
has made a great contribution here: the nonfactorizable
corrections to the naive factorizations of both the color-allowed
and color-suppressed modes can be predicted, and the latter is 
found to be very important.

\section{CONCLUSION}

In this paper we have analyzed the two-body charmed nonleptonic 
decays $B\to D^{(*)}M$ with $M=\pi$, $\rho$, and $\omega$ in the
PQCD approach. This framework is based on $k_T$ factorization
theorem, which is free of end-point singularities and
gauge-invariant. $k_T$ factorization theorem is more appropriate,
when the end-point region of a momentum fraction is important, and
collnear factorization theorem breaks down. By including the
transverse degrees of freedom of partons in the evaluation of a
hard kernel, and the Sudakov factors from $k_T$ and
threshold resummations, the virtual particles remain sufficiently
off-shell, and the end-point singularities do not exist. We have
explained that there is no conflict between LCSR with the 
soft-dominance picture and PQCD with the hard-dominance picture,
since the soft contributions refer to the different quantities in
the two theoretical frameworks. 

%When applying this approach to two-body charmless decays, we have
%predicted large CP asymmetries and dynamical penguin enhancement,
%which are in agreement with the $B\to PP$ ($\pi\pi$, $K\pi$)
%and $B\to VP$ ($\phi K$, $K^*\pi$) data, respectively.

The derivation of the factorization formulas for the $B\to
D^{(*)}M$ decay amplitudes follow the power counting rules
constructed in our previous work on the $B\to D^{(*)}$ transition
form factors. Under the hierachy $m_B\gg
m_{D^{(*)}}\gg\bar\Lambda$, the $B$ and $D^{(*)}$ meson wave
functions exhibit a peak at the momentum fractions around
$\bar\Lambda/m_B$ and $\bar\Lambda/m_{D^{(*)}}$, respectively. Up
to leading power in $m_{D^{(*)}}/m_B$ and in
$\bar\Lambda/m_{D^{(*)}}$, only a single $B$ meson wave function
and a single $D^{(*)}$ meson wave function are involved. The
factorization formulas then become simpler than those
for the charmless decays. Moreover, the factorization formulas for
all the $B\to D^{(*)}M$ modes are identical, except the
appropriate substitution of the masses, the decay constants, and
the meson distribution amplitudes. We emphasize that there is no 
arbitrary parameter in our analysis (there are in QCDF), though 
all universal inputs are not yet known precisely. The meson wave 
functions have been determined either from the semileptonic data 
or from LCSR.

Being free from the end-point singularities, all topologies of
decay amplitudes are calculable in PQCD, including the
color-suppressed nonfactorizable one. This amplitude can not be
computed in QCDF based on the collinear factorization theorem due to
the existence of the end-point singularities for a massive $c$
quark. We have observed in PQCD that this amplitude, not suppressed 
by the Wilson coefficient (proportional to $C_2/N_c$), is comparable
to the dominant color-allowed factorizable amplitude. It generates
a large strong phase from the non-pinched singularity of the hard
kernel, which is crucial for explaining the observed $B\to
D^{(*)}M$ branching ratios. The other topologies are less
important: the color-allowed nonfactorizable
contribution is negligible because of the pair cancellation and
the small Wilson coefficient $C_1/N_c$. The color-suppressed
factorizable amplitude with the small Wilson coefficient 
$a_2=C_1+C_2/N_c$ is also negligible. The annihilation amplitudes 
are small, since they come from the tree operators.

All our predictions are consistent with the existing measurements. 
For those without data, such as the $B\to D^{*0}\rho^0$, 
$D^{*0}\omega$ modes, our predictions can be confronted with 
future measurement. As stated before, we have predicted the large 
strong phases from the scalar-penguin annihilation amplitudes, which 
are required by the large CP asymmetries observed in two-body charmless 
decays. The success in predicting the storng phases from the 
color-suppressed nonfactorizable amplitudes for the two-body 
charmed decays further supports the $k_T$ factorization theorem. The 
conclusion drawn in this work is that the short-distance strong phase 
is already sufficient to account for the $B\to D^{(*)}M$ data. 
Certainly, there is still room for long-distance strong phases 
from final-state interaction. For the application of the PQCD 
approach to other charmed decays, such as $B\to D_s^{(*)} K$ and 
$B\to D^{(*)}f_0$, refer to \cite{LL03} and \cite{Chen03}, 
respectively.

\section*{Acknowledgments}
The authors are grateful to the organizers of Summer Institute
2002 at Fuji-Yoshida, Japan, where part of this work was done, for
warm hospitality. This work was supported by the Japan Society for
the Promotion of Science (Y.Y.K.), by Grant-in Aid for Scientific
Research from the Japan Society for the Promotion of Science under
the Grant No. 11640265 (T.K.), by the National Science Council of
R.O.C. under the Grant No. NSC-91-2112-M-001-053 (H-n.L.), by the
National Science Foundation of China under Grant Nos. 90103013 and
10135060 (C.D.L.), and by Ministry of Education, Science and
Culture, Japan (A.I.S.).

\appendix

%%%%%%%%%%%%%%%%%%%%%%%%  Appendix-A %%%%%%%%%%%%%%%%%%%%%%%%%%%%%%%%%
\section{FACTORIZATION FORMULAS FOR $B \to D\pi$ }

In this Appendix we present the factorization formulas for the $B
\to D\pi$ decay amplitudes. We choose the $B$ meson, $D$ meson,
and pion momenta in the light-cone coordinate as,
\begin{equation}
P_1 = \frac{m_B}{\sqrt{2}}(1,1,{\bf 0}_T)\;, \ \
P_2 = \frac{m_B}{\sqrt{2}}(1,r^2,{\bf 0}_T)\;, \ \
P_3 = \frac{m_B}{\sqrt{2}}(0,1-r^2,{\bf 0}_T)\;,
\end{equation}
respectively, with $r = m_{D}/m_B$ being defined before. The
fractional momenta of the light valence quarks in the $B$ meson,
$D$ meson and the pion are
\begin{eqnarray}
  k_1 &=& x_1\frac{m_B}{\sqrt{2}}(1,0,{\bf 0}_T) + {\bf k}_{1T}
\mbox{\ \  for\ \  $\xi_{\rm int}$, $M_{\rm int}$,}
\nonumber\\
  k_1 &=& x_1\frac{m_B}{\sqrt{2}}(0,1,{\bf 0}_T) + {\bf k}_{1T}
\mbox{\ \  for\ \  others,}
\nonumber\\
k_2 &=& x_2\frac{m_B}{\sqrt{2}}(1,0,{\bf 0}_T)+{\bf k}_{2T}\;,
\nonumber\\
k_3 &=& x_3\frac{m_B}{\sqrt{2}}(0,1-r^2,{\bf 0}_T)+{\bf k}_{3T}\;,
\label{k123}
\end{eqnarray}
respectively. Which longitudinal component of $k_1$, $k_1^+$
or $k_1^-$, is relevant depends on the final-state meson the
hard gluon attaches. That is, it is selcted by the inner
product $k_1\cdot k_3$ or $k_1\cdot k_2$.

The factorizable amplitudes $\xi_{\rm ext}$, $\xi_{\rm int}$ and
$\xi_{\rm exc}$ are written as
\begin{eqnarray}
\xi_{\rm ext} &=&16\pi C_F\sqrt{r}m_B^2
\int_{0}^{1}d x_{1}d x_{2}\int_{0}^{1/\Lambda} b_1d b_1 b_2d b_2
\phi_B(x_1,b_1)\phi_{D}(x_2)
\nonumber \\
& &\times
\left[ E_e(t_e^{(1)})h(x_1,x_2,b_1,b_2)S_t(x_2)
+ rE_e(t_e^{(2)})h(x_2,x_1,b_2,b_1)S_t(x_1)\right]\;,
\label{ext}\\
\xi_{\rm int}&=&16\pi C_F\sqrt{r}m_B^2
\int_0^1 dx_1dx_3\int_0^{1/\Lambda}b_1db_1b_3db_3
\phi_B(x_1,b_1)
\nonumber \\
& &\times \{[(1+x_3) \phi_\pi(x_3)
+ r_0 (1- 2x_3) (\phi_\pi^p(x_3)+ \phi_\pi^t(x_3))]
\nonumber\\
& &\times E_i(t_i^{(1)})h(x_1,x_3(1-r^2),b_1,b_3)S_t(x_3)
\nonumber \\
& &\ \ + 2r_0 \phi_\pi^p(x_3) E_i(t_i^{(2)})
h(x_3,x_1(1-r^2),b_3,b_1)S_t(x_1)\}\;,
\label{int} \\\
\xi_{\rm exc}&=&16\pi C_F\sqrt{r}m_B^2
\int_0^1 dx_2dx_3\int_0^{1/\Lambda}b_2db_2b_3db_3
\phi_{D}(x_2)
\nonumber \\
& &\times \left[-x_3\phi_\pi(x_3)
%- r r_0 (1+ 2x_3)\phi_\pi^p (x_3)
%\nonumber \\
%& &\ \ \ \ \ \
%+ r r_0 (1-2x_3)\phi_\pi^t  (x_3)]
E_a(t_a^{(1)}) h_a(x_2,x_3(1-r^2),b_2,b_3)S_t(x_3)\right.
\nonumber \\
& &\ \ \  \left.+x_2\phi_\pi(x_3)
% + 2rr_0 (1 + x_2)\phi_\pi^p(x_3)]
E_a(t_a^{(1)}) h_a(x_3,x_2(1-r^2),b_3,b_2)S_t(x_2)\right]\;,
\label{exc}
\end{eqnarray}
with the mass ratio $r_0 \equiv m_0/m_B$, the
evolution factors,
\begin{eqnarray}
E_e(t)&=&\alpha_s(t)a_1(t)\exp[-S_B(t)-S_{D}(t)]\;,
\nonumber \\
E_i(t)&=&\alpha_s(t)a_2(t)\exp[-S_B(t)-S_\pi(t)]\;,
\nonumber \\
E_a(t)&=&\alpha_s(t)a_2(t)\exp[-S_{D}(t)-S_\pi(t)]\;,
\end{eqnarray}
and the Wilson coefficients,
\begin{equation}
a_1=C_2+\frac{C_1}{N_c}\;,\;\;\;\;
a_2=C_1+\frac{C_2}{N_c}\;.
\end{equation}
Note that $C_1 =0$ and $C_2 = 1$ at tree level in our convention.
The explicit expressions of the Sudakov factors $\exp[-S_B(t)]$,
$\exp[-S_{D}(t)]$ and $\exp[-S_\pi(t)]$ from $k_T$ resummation are
referred to \cite {TLS2,TLS}.
%given by
%\begin{eqnarray}
%\exp[-S_B(\mu)]&=&\exp\left[-s(x_1P_1^-,b)-2\int_{1/b}^\mu
%\frac{d{\bar\mu}}{\bar\mu}\gamma(\alpha_s({\bar\mu}))\right]\;,
%\nonumber \\
%\exp[-S_{D^{(*)}}(\mu)]&=&\exp\left[-s(x_2P_2^+,b)-2\int_{1/b}^\mu
%\frac{d{\bar\mu}}{\bar\mu}\gamma(\alpha_s({\bar\mu}))\right]\;,
%\nonumber \\
%\exp[-S_\pi(\mu)]&=&\exp\left[-s(x_3P_3^-,b)-s((1-x_3)P_3^-,b)
%-2\int_{1/b}^\mu
%\frac{d{\bar\mu}}{\bar\mu}\gamma(\alpha_s({\bar\mu}))\right]\;.
%\label{ktd}
%\end{eqnarray}

The functions $h$'s, obtained from Figs.~4(a) and 4(b), Figs.~5(a)
and 5(b), and Figs.~6(a) and 6(b), are given by
\begin{eqnarray}
h(x_1,x_2,b_1,b_2)&=&K_{0}\left(\sqrt{x_1x_2}m_Bb_1\right)
\nonumber \\
& &\times \left[\theta(b_1-b_2)K_0\left(\sqrt{x_2}m_B
b_1\right)I_0\left(\sqrt{x_2}m_Bb_2\right)\right.
\nonumber \\
& &\left.+\theta(b_2-b_1)K_0\left(\sqrt{x_2}m_Bb_2\right)
I_0\left(\sqrt{x_2}m_Bb_1\right)\right]\;,
\label{dh}\\
h_a(x_2,x_3,b_2,b_3)&=&\left(i\frac{\pi}{2}\right)^2
H_0^{(1)}\left(\sqrt{x_2x_3}m_Bb_2\right)
\nonumber \\
& &\times\left[\theta(b_2-b_3)
H_0^{(1)}\left(\sqrt{x_3}m_Bb_2\right)
J_0\left(\sqrt{x_3}m_Bb_3\right)\right.
\nonumber \\
& &\left.+\theta(b_3-b_2)H_0^{(1)}\left(\sqrt{x_3}m_Bb_3\right)
J_0\left(\sqrt{x_3}m_Bb_2\right)\right]\;.
\end{eqnarray}
The hard scales $t$ are chosen as
\begin{eqnarray}
& &t_e^{(1)}={\rm max}(\sqrt{x_2}m_B,1/b_1,1/b_2)\;, \;\;\;
t_e^{(2)}={\rm max}(\sqrt{x_1}m_B,1/b_1,1/b_2)\;,
\nonumber \\
& &t_i^{(1)}={\rm max}(\sqrt{x_3(1-r^2)}m_B,1/b_1,1/b_3)\;, \;\;\;
t_i^{(2)}={\rm max}(\sqrt{x_1(1-r^2)}m_B,1/b_1,1/b_3)\;,
\nonumber\\
& &t_a^{(1)}={\rm max}(\sqrt{x_3(1-r^2)}m_B,1/b_2,1/b_3)\;,\;\;\;
t_a^{(2)}={\rm max}(\sqrt{x_2(1-r^2)}m_B,1/b_2,1/b_3)\;.
\end{eqnarray}

For the nonfactorizable amplitudes, the factorization formulas involve
the kinematic variables of all the three mesons. Their expressions are
\begin{eqnarray}
{\cal M}_{\rm ext}&=& 32\pi\sqrt{2N} C_F\sqrt{r}m_B^2
\int_0^1 [dx]\int_0^{1/\Lambda}
b_1 db_1 b_3 db_3
\phi_B(x_1,b_1)\phi_{D}(x_2)\phi_\pi(x_3)
\nonumber \\
& &\times \left[x_3E_b(t_b^{(1)})h^{(1)}_b(x_i,b_i)
-(1-x_3+x_2)E_b(t_b^{(1)})h^{(2)}_b(x_i,b_i) \right]\;,
\label{mb}\\
{\cal M}_{\rm int}&=& 32\pi\sqrt{2N} C_F\sqrt{r}m_B^2
\int_0^1 [dx]\int_0^{1/\Lambda}b_1 db_1 b_2 db_2
\phi_B(x_1,b_1)\phi_D(x_2)
\nonumber \\
& &\times   \left[(-x_2-x_3)\phi_\pi(x_3)
% +r_0x_3(\phi_\pi^p(x_3) +\phi_\pi^t(x_3))]
E_d(t_d^{(1)})h^{(1)}_d(x_i,b_i)
+(1-x_2)\phi_\pi(x_3)
%-r_0x_3(\phi_\pi^p(x_3) -\phi_\pi^t(x_3)) ]
E_d(t_d^{(2)})h^{(2)}_d(x_i,b_i)\right]\;,
\label{md}\\
{\cal M}_{\rm exc}&=& 32 \pi\sqrt{2N} C_F\sqrt{r}m_B^2
\int_0^1 [dx]\int_0^{1/\Lambda}b_1 db_1 b_2 db_2
\phi_B(x_1,b_1)\phi_{D}(x_2)
\nonumber \\
& &\times \left[x_3\phi_\pi(x_3)
%+rr_0( x_2 + x_3)\phi_\pi^p(x_3)
%\nonumber \\
%& &\ \ \ \ \ \ \
%-rr_0(x_2 - x_3)\phi_\pi^t(x_3)]
E_f(t_f^{(1)})h^{(1)}_f(x_i,b_i)
-x_2\phi_\pi(x_3)
%+rr_0(2 + x_2 + x_3)\phi_\pi^p(x_3)
%\nonumber \\
%& &\ \ \ \ \ \ \
%+ rr_0(x_2 - x_3)\phi_\pi^t(x_3)]
E_f(t_f^{(2)})h^{(2)}_f(x_i,b_i) \right]\;,
\label{mf}
\end{eqnarray}
from Figs.~4(c) and 4(d), Figs.~5(c) and 5(d), and Figs.~6(c) and
6(d), respectively, with the definition $[dx]\equiv dx_1dx_2dx_3$.
The evolution factors are given by
\begin{eqnarray}
E_b(t)&=&\alpha_s(t)\frac{C_1(t)}{N}\exp[-S(t)|_{b_2=b_1}]\;,
\nonumber \\
E_d(t)&=&\alpha_s(t)\frac{C_2(t)}{N}\exp[-S(t)|_{b_3=b_1}]\;,
\nonumber \\
E_f(t)&=&\alpha_s(t)\frac{C_2(t)}{N}\exp[-S(t)|_{b_3=b_2}]\;.
\end{eqnarray}
with the Sudakov exponent $S=S_B+S_{D}+S_\pi$.

The functions $h^{(j)}$, $j=1$ and 2, appearing in
Eqs.~(\ref{mb})-(\ref{mf}), are written as
\begin{eqnarray}
\everymath{\displaystyle}
h^{(j)}_b&=& \left[\theta(b_1-b_3)K_0\left(Bm_B
b_1\right)I_0\left(Bm_Bb_3\right)\right. \nonumber \\
& &\quad \left.
+\theta(b_3-b_1)K_0\left(Bm_B b_3\right)
I_0\left(Bm_B b_1\right)\right]
\nonumber \\
&  & \times \left( \begin{array}{cc}
 K_{0}(B_{j}m_Bb_{3}) &  \mbox{for $B^2_{j} \geq 0$}  \\
 \frac{i\pi}{2} H_{0}^{(1)}(\sqrt{|B_{j}^2|}m_Bb_{3})  &
 \mbox{for $B^2_{j} \leq 0$}
  \end{array} \right)\;,
\\
\everymath{\displaystyle}
h^{(j)}_d&=& \left[\theta(b_1-b_2)K_0\left(Dm_B
b_1\right)I_0\left(Dm_Bb_2\right)\right. \nonumber \\
& &\quad \left.
+\theta(b_2-b_1)K_0\left(Dm_B b_2\right)
I_0\left(Dm_B b_1\right)\right]
 \nonumber \\
&  & \times \left( \begin{array}{cc}
 K_{0}(D_{j}m_Bb_{2}) &  \mbox{for $D^2_{j} \geq 0$}  \\
 \frac{i\pi}{2} H_{0}^{(1)}(\sqrt{|D_{j}^2|}m_Bb_{2})  &
 \mbox{for $D^2_{j} \leq 0$}
  \end{array} \right)\;,
\label{hjd}\\
\everymath{\displaystyle}
h^{(j)}_f&=& i\frac{\pi}{2}
\left[\theta(b_1-b_2)H_0^{(1)}\left(Fm_B
b_1\right)J_0\left(Fm_Bb_2\right)\right. \nonumber \\
& &\quad\left.
+\theta(b_2-b_1)H_0^{(1)}\left(Fm_B b_2\right)
J_0\left(Fm_B b_1\right)\right]\;  \nonumber \\
&  & \times \left( \begin{array}{cc}
 K_{0}(F_{j}m_Bb_{1}) &  \mbox{for $F^2_{j} \geq 0$}  \\
 \frac{i\pi}{2} H_{0}^{(1)}(\sqrt{|F_{j}^2|}m_Bb_{1})  &
 \mbox{for $F^2_{j} \leq 0$}
  \end{array} \right)\;,
\end{eqnarray}
with the variables
\begin{eqnarray}
B^{2}&=&x_{1}x_{2}\;,
\nonumber \\
B_{1}^{2}&=&x_{1}x_{2}-x_2x_{3}(1-r^{2})\;,
\nonumber \\
B_{2}^{2}&=& x_{1}x_{2}-x_2(1-x_{3})(1-r^{2})\;,
\nonumber \\
D^{2}&=&x_{1}x_{3}(1-r^{2})\;,
\nonumber \\
D_{1}^{2}&=&F_1^2=(x_{1}-x_{2})x_{3}(1-r^{2})\;,
\nonumber \\
D_{2}^{2}&=&(x_{1}+x_{2})r^{2}-(1-x_{1}-x_{2})x_{3}(1-r^{2})\;,
\nonumber \\
F^{2}&=&x_{2}x_{3}(1-r^{2})\;,
\nonumber \\
F_{2}^{2}&=&x_{1}+x_{2}+(1-x_{1}-x_{2})x_{3}(1-r^{2})\;.
\label{mis}
\end{eqnarray}
There is an ambiguity in defining a light-cone $B$ meson wave
function for the nonfactorizable amplitude ${\cal M}_{\rm exc}$,
since both the components $k_1^+$ and $k_1^-$ contribute through
the inner products $k_1\cdot k_2$ and $k_1\cdot k_3$ in the
denominators of the virtual particle propagators. However, a
careful examination of the factorization formula shows that the
dominant region is the one with $k_2\sim O(\bar\Lambda)$ and
$k_3\sim O(m_B)$ at leading twist. Hence, we drop the term
$k_1\cdot k_2$. The scales $t^{(j)}$ are chosen as
\begin{eqnarray}
t_b^{(j)}&=&{\rm max}(Bm_B,\sqrt{|B_j^2|}m_B,1/b_1,1/b_3)\;,
\nonumber \\
t_d^{(j)}&=&{\rm max}(Dm_B,\sqrt{|D_j^2|}m_B,1/b_1,1/b_2)\;,
\nonumber \\
t_f^{(j)}&=&{\rm max}(Fm_B,\sqrt{|F_j^2|}m_B,1/b_1,1/b_2)\;.
\end{eqnarray}

We explain that the factorization formulas presented above are
indeed of leading power under the power counting rules in
\cite{TLS2}. The factorizable amplitudes are as shown in
\cite{TLS2,CKL}. For the nonfactorizable amplitudes, the terms
proportional to $x_3$ and to $1-x_3$ in ${\cal M}_{\rm ext}$
cancel each other roughly. This cancellation can be understood by
means of the corresponding expression in collinear factorization
theorem: the first and second terms in ${\cal M}_{\rm ext}$ are
proportional to
\begin{eqnarray}
-\frac{x_3}{x_1x_2^2x_3}\;,\;\;\;\;
\frac{1-x_3+x_2}{x_1x_2^2(1-x_3)}\;.
\end{eqnarray}
For simplicity, $x_1$ has been suppressed, when it appears in the sum
together with $x_2$ or $x_3$. It is found that the first ratio
cancels the $1-x_3$ term in the second ratio. That is, the $x_2$ term
is in fact leading and not negligible.
For a similar reason, the $-x_2$ term in ${\cal M}_{\rm int}$
cancels the $1-x_2$ term. Hence, the $-x_3$ term is leading.
If one drops $-x_2$ in ${\cal M}_{\rm int}$, the above cancellation
disappears, and a fake leading term will be introduced.

The pion and $\rho$ meson distribution amplitudes have been
derived in \cite{PB1,PB2}:
\begin{eqnarray}
\phi_\pi(x)&=&\frac{3f_\pi}{\sqrt{2N_c}} x(1-x)
\left[1+0.44C_2^{3/2}(2x-1)+0.25C_4^{3/2}(2x-1)\right]\;,
\label{pioa}\\
\phi_\pi^p(x)&=&\frac{f_\pi}{2\sqrt{2N_c}}
\left[1+0.43C_2^{1/2}(2x-1)+0.09C_4^{1/2}(2x-1)\right]\;,
\label{piob}\\
\phi_\pi^t(x)&=&\frac{f_\pi}{2\sqrt{2N_c}} (1-2x)
\left[1+0.55(10x^2-10x+1)\right]\;,
\label{pioc}\\
\phi_\rho(x)&=&\frac{3f_\rho}{\sqrt{2N_c}} x(1-x)\left[1+
0.18C_2^{3/2}(2x-1)\right]\;,
\label{pwr}\\
\phi_{\rho}^t(x)&=&\frac{f^T_{\rho}}{2\sqrt{2N_c}}
\left\{3(2x-1)^2+0.3(2x-1)^2[5(2x-1)^2-3]\right.
\nonumber \\
& &\left.+0.21[3-30(2x-1)^2+35(2x-1)^4]\right\}\;,
\label{pwt}\\
\phi_{\rho}^s(x) &=&\frac{3f_\rho^T}{2\sqrt{2N_{c}}}
(1-2x)\left[1+0.76(10x^2-10x+1)\right]\;,
\label{pws}\\
\phi_\rho^T(x)&=&\frac{3f_\rho^T}{\sqrt{2N_c}} x(1-x)\left[1+
0.2C_2^{3/2}(2x-1)\right]\;,
\label{pwft}\\
\phi_{\rho}^v(x)&=&\frac{f_{\rho}}{2\sqrt{2N_c}}
\bigg\{\frac{3}{4}[1+(2x-1)^2]+0.24[3(2x-1)^2-1]
\nonumber \\
& &+0.12[3-30(2x-1)^2+35(2x-1)^4]\bigg\}\;,
\label{pwv}\\
\phi_{\rho}^a(x) &=&\frac{3f_\rho}{4\sqrt{2N_{c}}}
(1-2x)\left[1+0.93(10x^2-10x+1)\right]\;,
\label{pwa}
\end{eqnarray}
with the Gegenbauer polynomials,
\begin{eqnarray}
& &C_2^{1/2}(t)=\frac{1}{2}(3t^2-1)\;,\;\;\;
C_4^{1/2}(t)=\frac{1}{8}(35 t^4 -30 t^2 +3)\;,
\nonumber\\
& &C_2^{3/2}(t)=\frac{3}{2}(5t^2-1)\;,\;\;\;
C_4^{3/2}(t)=\frac{15}{8}(21 t^4 -14 t^2 +1) \;.
\end{eqnarray}
We shall assume that the $\omega$ meson wave functions
are identical to the $\rho$ meson ones in this work.

\newpage
%%%%%%%%%%%%%%%%%%%%%      References  %%%%%%%%%%%%%%%%%%%%%%%%%%%%%%

%%%%%%%%%%%%%%%%%%%%%%%%    TABLES     %%%%%%%%%%%%%%%%%%%%%%%%%%%%%%%%%

\newpage
%%%%%%%%%%%%%%%%%%%%%  Table-1 %%%%%%%%%%%%%%%%%%%%%%%%%%%
\begin{table}[htb]
\begin{center}
\begin{tabular}{|l| l| c| } \hline
{Decay mode} &
 Belle \cite{BelleC} & CLEO \cite{CLEOC} \\ \hline
 $\bar B^0\to D^0\pi^0$ &  $3.1\pm0.4\pm0.5$ &
 $2.74^{+0.36}_{-0.32}\pm 0.55$ \\
 $\bar B^0\to D^{*0}\pi^0$ &  $2.7^{+0.8+0.5}_{-0.7-0.6}$ &
 $2.20^{+0.59}_{-0.52}\pm0.79$ \\
 $\bar B^0\to D^0\eta$ & $1.4^{+0.5}_{-0.4}\pm 0.3$ & \\
 $\bar B^0\to D^{*0}\eta$ & $2.0^{+0.9}_{-0.8}\pm0.4$ & \\
 $\bar B^0\to D^0\omega$ &  $1.8\pm0.5^{+0.4}_{-0.3}$ & \\
 $\bar B^0\to D^{*0}\omega$ &  $3.1^{+1.3}_{-1.1}\pm0.8$ & \\ \hline
\end{tabular}
\caption{Data (in units of $10^{-4}$) of the
$\bar B^0\to D^{(*)0}M^0$ $(X=\pi,\eta,\omega)$ branching ratios.}
\label{dpda}
\end{center}
\end{table}
%%%%%%%%%%%%%%%%%%%%%%%%%%%%%%%%%%%%%%%%%%%%%%%%%%%%%%%%%%%%%%%
%%%%%%%%%%%%%%%%%%%  Table-2  %%%%%%%%%%%%%%%%%%%%%%%
\begin{table}[htb]
\begin{center}
\begin{tabular}{|l| c| c|c| }
\hline Amplitudes & $C_D=0.6$ & $C_D=0.8$ & $C_D=1.0$  \\
\hline
$f_\pi\xi_{\rm ext}$ & $6.90$ & $7.46$ & $8.01$ \\
$f_D\xi_{\rm int}$ & $-1.44$ & $-1.44$ & $-1.44$ \\
$f_B\xi_{\rm exc}$ &$-0.01-0.03i$ &$-0.02-0.03i$ & $-0.02-0.03i$\\
${\cal M}_{\rm ext}$ &$-0.24+0.57i$ &$-0.25+0.60i$ &$-0.27+0.65i$ \\
${\cal M}_{\rm int}$ &$3.34-3.02i$ &$3.22-3.07i$ &$3.10-3.12i$ \\
${\cal M}_{\rm exc}$ &$-0.26-0.89i$ &$-0.31-0.95i$ &$-0.37-1.02i$ \\
\hline
\end{tabular}
\caption{Predicted $B\to D\pi$ decay amplitudes in units of $10^{-2}$
GeV.}
\label{dpt}
\end{center}
\end{table}
%%%%%%%%%%%%%%%%%%%%%%%%%%%%%%%%%%%%%%%%%%%%%%%%%%%%%%%%%%%%%%

%%%%%%%%%%%%%%%%%%%  Table-3  %%%%%%%%%%%%%%%%%%%%%%%
\begin{table}
\begin{center}
\begin{tabular}{|l| c| c|c|c| }
\hline
Quantities & $C_D=0.6$ & $C_D=0.8$ & $C_D=1.0$ & Data \\ \hline
$A_1$ &$6.39-0.35i$ &$6.88-0.38i$ &$7.35-0.40i$ &\\
$A_2$ &$-1.53+1.48i$ &$-1.49+1.48i$ &$-1.45+1.45i$ &\\
$A_3$ &$8.56-2.45i$ &$8.99-2.47i$ &$9.40-2.46i$ &\\
\hline
$B(\bar B^0\to D^+\pi^-)$ & 2.37 & 2.74 & 3.13 & $3.0\pm 0.4$ \\
$B(\bar B^0\to D^0\pi^0)$ & 0.26 & 0.25 & 0.24 & $0.29 \pm 0.05$\\
$B(B^-\to D^0\pi^-)$ & 4.96 & 5.43 & 5.91 & $5.3\pm 0.5$ \\
\hline
$|a_2/a_1|$ (w/o anni.) & 0.47(0.51) & 0.43(0.46) & 0.39(0.42) &\\
$Arg(a_2/a_1)$ (w/o anni.) & $-42.5^o$($-61.5^o$ )
& $-41.6^o$ ($-63.5^o$) & $-41.9^o$($-65.3^o$) &\\
\hline
\end{tabular}
\caption{Predicted $B\to D\pi$ decay amplitudes in units of
$10^{-2}$ GeV, branching ratios in units of $10^{-3}$,
$|a_2/a_1|$, and relative angle $Arg(a_2/a_1)$ in units of
degree.} \label{dp}
\end{center}
\end{table}
%%%%%%%%%%%%%%%%%%%%%%%%%%%%%%%%%%%%%%%%%%%%%%%%%%%%%%%%%%%%%%
%%%%%%%%%%%%%%%%%%%%%%%  Table-4  %%%%%%%%%%%%%%%%%%%%%%
\begin{table}
\begin{center}
\begin{tabular}{|l| c| c|c|c| }
\hline
Quantities & $C_{D^*}=0.5$ & $C_{D^*}=0.7$ & $C_{D^*}=0.9$ & Data \\
\hline
$A_1$ &$6.32-0.42i$ &$6.81-0.45i$ &$7.30-0.49i$ &\\
$A_2$ &$-1.65+1.61i$ &$-1.62+1.59i$ &$-1.59+1.57i$ &\\
$A_3$ &$8.65-2.69i$ &$9.10-2.70i$ &$9.55-2.70i$ &\\
\hline
$B(\bar B^0\to D^{*+}\pi^-)$ & 2.16 & 2.51 & 2.88 & $2.76\pm 0.21$ \\
$B(\bar B^0\to D^{*0}\pi^0)$ & 0.29 & 0.28 & 0.27 & $0.25 \pm 0.07$\\
$B(B^-\to D^{*0}\pi^-)$ & 4.79 & 5.26 & 5.75 & $4.60\pm 0.40$ \\
\hline
$|a_2/a_1|$ (w/o anni.) & 0.52 (0.55) & 0.47 (0.50) & 0.43 (0.47)&\\
$Arg(a_2/a_1)$ (w/o anni.) & $-40.5^o$ ($-61.4^o$)
& $-40.7^o$ ($-63.1^o$) & $-40.8^o$ ($-64.8^o$)&\\
\hline
\end{tabular}
\caption{Predicted $B\to D^*\pi$ decay amplitudes in units of
$10^{-2}$ GeV, branching ratios in units of $10^{-3}$,
$|a_2/a_1|$, and relative angle  $Arg(a_2/a_1)$ in units of
degree.} \label{dsp}
\end{center}
\end{table}
%%%%%%%%%%%%%%%%%%%%%%%%%%%%%%%%%%%%%%%%%%%%%%%%%%%%%%%%
%%%%%%%%%%%%%%%%%%%%   Table-5  %%%%%%%%%%%%%%%%%%%%%%%%%%%%%%%
\begin{table}
\begin{center}
\begin{tabular}{|l| c| c|c|c| }
\hline
Branching ratios & $C_D=0.6$ & $C_D=0.8$ & $C_D=1.0$ & Data \\ \hline
$B(\bar B^0\to D^{+}\rho^-)$ & 4.10 & 4.72 & 5.38 & $8.0\pm 1.4$ \\
$B(\bar B^0\to D^{0}\rho^0)$ & 0.17 & 0.17 & 0.17 & $0.29\pm 0.11$\\
$B(B^-\to D^{0}\rho^-)$ & 7.26 & 8.15 & 9.09 & $13.4\pm 1.8$ \\
$B(\bar B^0\to D^{0}\omega)$ & 0.50 & 0.54 & 0.56 & $0.18\pm 0.06$\\
\hline
Branching ratios & $C_{D^{*}}=0.5$ & $C_{D^{*}}=0.7$
& $C_{D^{*}}=0.9$ & Data \\ \hline
$B(\bar B^0\to D^{*+}\rho^-)$ & 5.32 & 6.16 & 7.08 & $7.3\pm 1.5$ \\
$B(\bar B^0\to D^{*0}\rho^0)$ & 0.18 & 0.18 & 0.19 & $<0.51$\\
$B(B^-\to D^{*0}\rho^-)$ & 9.14 & 10.32 & 11.60 & $15.5\pm 3.1$ \\
$B(\bar B^0\to D^{*0}\omega)$ & 0.49 & 0.53 & 0.58 & $<0.74$\\
\hline
\end{tabular}
\caption{Predicted $B\to D^{(*)}\rho(\omega)$ branching ratios in
units of $10^{-3}$.}
\label{dr}
\end{center}
\end{table}
%%%%%%%%%%%%%%%%%%%%%%%%%%%%%%%%%%%%%%%%%%%%%%%%%%%%%%%%%%%%%%
%%%%%%%%%%%%%%%%%%%%%%%%%%%%%%%%%%%%%%%%%%%%%%%%%%%%%%%%%%%%%%%%%%%%%%%%%
\end{document}